\documentclass[aps,pra]{revtex4}

\usepackage{amssymb}
\usepackage{graphicx}
\usepackage{dcolumn}
\usepackage{bm}
\usepackage{amsmath}
\usepackage{epsfig}

\begin{document}

\title{Thermodynamic analysis of quantum light amplification}

\author{E. Boukobza}
\author{D.J. Tannor}
\affiliation{Department of Chemical Physics,\\
Weizmann Institute of Science, Rehovot 76100, Israel}

\begin{abstract}

Thermodynamics of a three-level maser was studied in the pioneering
work of Scovil and Schulz-DuBois [Phys. Rev. Lett. {\bf2}, 262
(1959)]. In this work we consider the same three-level model, but
treat both the matter and light quantum mechanically.  Specifically,
we analyze an extended (three-level) dissipative Jaynes-Cummings
model (ED-JCM) within the framework of a quantum heat engine, using
novel formulas for heat flux and power in bipartite systems
introduced in our previous work [E. Boukobza and D. J. Tannor, PRA
(in press)]. Amplification of the selected cavity mode occurs even
in this simple model, as seen by a positive steady state power.
However, initial field coherence is lost, as seen by the decaying
off-diagonal field density matrix elements, and by the Husimi-Kano Q
function. We show that after an initial transient time the field's
entropy rises linearly during the operation of the engine, which we
attribute to the dissipative nature of the evolution and not to
matter-field entanglement. We show that the second law of
thermodynamics is satisfied in two formulations (Clausius, Carnot)
and that the efficiency of the ED-JCM heat engine agrees with that
defined intuitively by Scovil and Schulz-DuBois.  Finally, we
compare the steady state heat flux and power of the fully quantum
model with the semiclassical counterpart of the ED-JCM, and derive
the engine efficiency formula of Scovil and Schulz-DuBois
\textit{analytically} from fundamental thermodynamic fluxes.

\end{abstract}

\maketitle

\section{Introduction}

Thermodynamics of quantum-optical systems has intrigued scientists
ever since masers and lasers were realized experimentally. Scovil
and Schulz-DuBois \cite{Scovil01} analyzed a three-level maser in
the framework of a heat engine. Based on a Boltzmann distribution of
atomic populations, they gave an intuitive definition of the
engine's efficiency, and showed it to be less than an or equal to
the Carnot efficiency. Using the concept of negative temperature
\cite{Purcell}, and motivated by Ramsey's \cite{Ramsey} work on
'spin temperature', Scovil and Schulz-DuBois \cite{Scovil02}
extended their analysis of three level systems to cases where the
reservoirs' temperature is negative, and introduced the concept of
negative efficiencies. Alicki studied a generic open quantum system
coupled to heat reservoirs, and under the influence of varying
external conditions (such as a time dependent field) \cite{Alicki}.
Alicki partitioned the energy of a quantum system into heat and work
using the time dependencies of the density and Hamiltonian
operators. Based on Alicki's definitions for heat and work, Kosloff
analyzed two coupled oscillators interacting with hot and cold
thermal reservoirs in the framework of a heat engine, and showed
that the engine's efficiency complies with the second law of
thermodynamics \cite{Kosloff01}. In later work, Geva and Kosloff
studied a three-level amplifier coupled to two heat reservoirs
\cite{Geva03} \cite{Geva04}. In their model the external field
influences the dissipative terms, and the second law of
thermodynamics is generally satisfied.

This paper is to some extent a continuation of the studies discussed
in the previous paragraph.  In contrast with previous work, in our
approach the matter and the radiation field are treated as a
bipartite system that is fully quantized, as opposed to a forced
unipartite system.  This treatment of the working medium (the
material system) and the work source (the radiation field) on an
equal footing requires some new thermodynamic developments, that we
adapt from \cite{Erez02}.  The general methodology is applied to an
extended dissipative Jaynes-Cummings model (ED-JCM), which consists
of a three-level material system coupled to two thermal heat baths
and a quantized cavity mode.  We show that this system provides a
simple model of light amplification, which can then be analyzed
using formulations of the first and second law of thermodynamics for
bipartite systems.  The heat flux and power calculated with this
model lead to an engine efficiency that is in quantitative agreement
with the efficiency formula intuitively defined by Scovil and
Schulz-DuBois. A semiclassical counterpart of the ED-JCM equations
is then presented and solved completely at steady state, giving the
efficiency formula of Scovil and Schultz-DuBois analytically from
fundamental thermodynamic fluxes.

This paper is arranged in the following manner. Section II is a
brief introduction to the thermodynamics of bipartite systems. In
Section III we define the ED-JCM master equation. In Section IV we
present numerical results for the ED-JCM model, showing that it acts
as a simple model for a quantum amplifier. In Section V we discuss
the entropic behavior of the full system and its individual
components, its behavior at steady state and the role of
entanglement. In Section VI we give a thermodynamical analysis of
the ED-JCM. We formulate the first law of thermodynamics in two
different ways. We then show that the second law of thermodynamics
is satisfied in two formulations (Clausius, Carnot), and that the
efficiency of the ED-JCM heat engine agrees with that defined
intuitively by Scovil and Schulz-DuBois. In Section VII we compare
the steady state heat flux and power of the fully quantum model with
a semiclassical version of the ED-JCM, and derive the engine
efficiency formula of Scovil and Schulz-DuBois \textit{analytically}
from fundamental thermodynamic fluxes. Section VIII concludes.

\section{Thermodynamics of bipartite systems}

A bipartite system is described by a density matrix of a
$C^{m}\otimes C^{n}$ Hilbert space. The partial density matrix of
one part is obtained by tracing over the other:
\begin{equation}
\rho_{A(B)}=\textrm{Tr}_{B(A)}\{\mbox{\boldmath$\rho_{AB}$}\}.
\end{equation}
The entropy of a quantum system is given by the von Neumann entropy
\cite{von Neumann}:
\begin{equation}
S=-k_{B}\textrm{Tr}\{\rho\ln\rho\}.
\end{equation}

The evolution of a bipartite system is given by the following master
equation:
\begin{equation}
\mbox{\boldmath$\dot{\rho}_{AB}$}=\mathcal{L}_{h}[\mbox{\boldmath$\rho_{AB}$}]+\mathcal{L}_{d}[\mbox{\boldmath$\rho_{AB}$}],\label{masterAB}
\end{equation}
where
$\mathcal{L}_{h}[\mbox{\boldmath$\rho_{AB}$}]=-\frac{i}{\hbar}[\mbox{\boldmath$H$},
\mbox{\boldmath$\rho_{AB}$}]$ is the Hamiltonian part of the
Lindblad super operator, and $\mathcal{L}_{d}[\rho_{AB}]$ is the
dissipative part of the Lindblad super operator. The bipartite time
independent Hamiltonian is given by:
\begin{equation}
\mbox{\boldmath$H$}=\mbox{\boldmath$H_{A}$}+\mbox{\boldmath$H_{B}$}+\mbox{\boldmath$V_{AB}$},\label{HAB}
\end{equation}
where $\mbox{\boldmath$H_{A}$}=H_{A}\otimes\openone_{B}$ and
$\mbox{\boldmath$H_{B}$}=\openone_{A}\otimes H_{B}$ are the
Hamiltonians of subsystems $A$ and $B$, and
$\mbox{\boldmath$V_{AB}$}$ is the coupling term between them. Here
and throughout the article, we use bold letters to signify operators
that have a tensor product structure.

Heat flux and power of the individual parts of the system are
defined by \cite{Erez02}:
\begin{eqnarray}
\dot{Q}_{A(B)}&\equiv&\textrm{Tr}\{\mathcal{L}_{d}[\mbox{\boldmath$\rho_{AB}$}]\mbox{\boldmath$H_{A(B)}$}\}\\\nonumber
P_{A(B)}&\equiv&-\frac{i}{\hbar}\textrm{Tr}\{\mbox{\boldmath$\rho_{AB}$}[\mbox{\boldmath$H_{A(B)}$},
\mbox{\boldmath$V_{AB}$}]\}.\label{QdotPAB}
\end{eqnarray}
The energy flux of the full system is due only to the dissipative
part of the Lindblad super operator:
\begin{equation}
\dot{E}_{AB}=\textrm{Tr}\{\mathcal{L}_{d}[\mbox{\boldmath$\rho_{AB}$}]\mbox{\boldmath$H$}\}.\label{EAB}
\end{equation}

\section{The ED-JCM master equation}

\begin{figure}[htb]
\begin{center}
\includegraphics{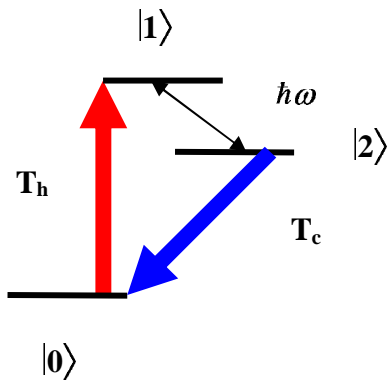}
\end{center}
\caption{\label{EDJCM}Three level system interacting with two heat
reservoirs (hot and cold) and a quantized cavity mode.}
\end{figure}

Consider a three-level system interacting resonantly with one
quantized cavity mode and two thermal photonic reservoirs as
depicted in Fig. \ref{EDJCM}.
The system is governed by the following master equation in the
interaction picture:
\begin{equation}
\mbox{\boldmath$\dot{\rho}_{mf}$}=\mathcal{L}_{h}[\mbox{\boldmath$\rho_{mf}$}]+\mathcal{L}_{dC}[\mbox{\boldmath$\rho_{mf}$}]+\mathcal{L}_{dH}[\mbox{\boldmath$\rho_{mf}$}].\label{MQE}
\end{equation}
The letters in the subscripts have the following significance:
$m$=matter, $f$=field, $d$=dissipative, $h$=Hamiltonian, $C$=cold,
$H$=hot. The Hamiltonian part of the Liouvillian is given by:
\begin{equation}
\mathcal{L}_{h}[\mbox{\boldmath$\rho_{mf}$}]=-\frac{i}{\hbar}[\mbox{\boldmath$V_{mf}$},\mbox{\boldmath$\rho_{mf}$}],
\label{eq:ixn_rep_h}
\end{equation}
where
\begin{equation}
\mbox{\boldmath$V_{mf}$}=\lambda(\sigma_{21}\otimes
a^{\dag}+\sigma_{21}^{\dag}\otimes a)
\end{equation}
is a resonant JCM type interaction Hamiltonian, $\lambda$ being the
matter-field coupling constant.
$\mathcal{L}_{dC}[\mbox{\boldmath$\rho_{mf}$}]$ and
$\mathcal{L}_{dH}[\mbox{\boldmath$\rho_{mf}$}]$ are the dissipative
cold and hot Lindblad super operators, respectively:
\begin{eqnarray}
\!\!\!\!\!\!\!\!\!\!\!\!\!\!\!\!\!\!\!\!\!\!\!\!\!\!\!\!\!\!\!\!\!\mathcal{L}_{dC}[\mbox{\boldmath$\rho_{mf}$}]&=&\Gamma_{02}\{(n_{02}\!+\!1)([\mbox{\boldmath$\sigma_{02}\rho_{mf}$},\mbox{\boldmath$\sigma_{02}^{\dag}$}]\!+\![\mbox{\boldmath$\sigma_{02}$},\!\mbox{\boldmath$\rho_{mf}\sigma_{02}^{\dag}$}])\!+\!
n_{02}([\mbox{\boldmath$\sigma_{02}^{\dag}\rho_{mf}$},\mbox{\boldmath$\sigma_{02}$}]\!+\![\mbox{\boldmath$\sigma_{02}^{\dag}$},\mbox{\boldmath$\rho_{mf}\sigma_{02}$}])\}\nonumber\\
\!\!\!\!\!\!\!\!\!\!\!\!\!\!\!\!\!\!\!\!\!\!\!\!\!\!\!\!\!\!\!\!\!\mathcal{L}_{dH}[\mbox{\boldmath$\rho_{mf}$}]&=&\Gamma_{01}\{(n_{01}\!+\!1)([\mbox{\boldmath$\sigma_{01}\rho_{mf}$},\mbox{\boldmath$\sigma_{01}^{\dag}$}]\!+\![\mbox{\boldmath$\sigma_{01}$},\mbox{\boldmath$\rho_{mf}\sigma_{01}^{\dag}$}])\!+\!
n_{01}([\mbox{\boldmath$\sigma_{01}^{\dag}\rho_{mf}$},\mbox{\boldmath$\sigma_{01}$}]\!+\![\mbox{\boldmath$\sigma_{01}^{\dag}$},\mbox{\boldmath$\rho_{mf}\sigma_{01}$}])\},\label{Ldch}
\end{eqnarray}
where $\Gamma_{02}$ and $\Gamma_{01}$ are the Weiskopf-Wigner decay
constant associated with the cold and hot reservoirs, respectively,
and $n_{02}$ and $n_{01}$ are the number of thermal photons in the
cold and hot reservoirs, respectively. Note that direct dissipation
occurs only through matter-reservoir coupling (the cold photonic
reservoir couples levels $|0\rangle$ and $|2\rangle$, the hot
photonic reservoir couples levels $|0\rangle$ and $|1\rangle$), and
is typically used to represent atomic decay in quantum optics
\cite{S&Z}. The matter creation and annihilation operators are in
tensor product form
$\mbox{\boldmath$\sigma_{ij}$}=\sigma_{ij}\otimes\openone_{f}$, and
their matrix form is given by:
\begin{equation*}
\sigma_{21}=\left(
\begin{array}{ccc}
0 & 0 & 0\\
0 & 0 & 0\\
0 & 1 & 0
\end{array}\right)\ \ \ \ \
\sigma_{01}=\left(
\begin{array}{ccc}
0 & 1 & 0\\
0 & 0 & 0\\
0 & 0 & 0
\end{array}\right)\ \ \ \ \
\sigma_{02}=\left(
\begin{array}{ccc}
0 & 0 & 1\\
0 & 0 & 0\\
0 & 0 & 0
\end{array}\right).
\end{equation*}
The reservoirs' temperature is given by:
\begin{equation}
T_{C(H)}=\frac{\hbar\omega_{C(H)}}{k_{B}\ln(1/n_{02(01)}+1)}\label{restemp},
\end{equation}
where $\omega_{C(H)}$ is the central frequency of the cold (hot)
reservoir. The ED-JCM master equation (equation \ref{MQE}) can be
obtained by summing the Hamiltonian contribution and the two
dissipative contributions. Alternatively, it can be derived for a
three-level system with a break in symmetry using the weak coupling
(to the reservoirs), Markovian, and Weiskopf-Wigner approximations
in a similar fashion to the simple JCM with master equation with
atomic damping which is derived in Appendix I.

The Hamiltonian (energy operator) of the full matter-field system is
given by:
\begin{equation}
\mbox{\boldmath$H$}=\mbox{\boldmath$H_{m}$}+\mbox{\boldmath$H_{f}$}+\mbox{\boldmath$V_{mf}$},\label{JCMH}
\end{equation}
where $\mbox{\boldmath$H_{m}$}=H_{m}\otimes\openone_{f};\
H_{m}=\hbar\sigma$ and $\mbox{\boldmath$H_{f}$}=\openone_{m}\otimes
H_{f};\ H_{f}=\hbar\omega_{f}a^{\dag}a$ are the matter and field
Hamiltonians, respectively, and $\mathbf{\sigma}$ is given by:
\begin{equation*}
\sigma=\left(
\begin{array}{ccc}
\omega_{0} & 0 & 0\\
0 & \omega_{1} & 0\\
0 & 0 & \omega_{2}
\end{array}\right).
\end{equation*}
Under matter-field resonance
($\omega_{m}=\omega_{1}-\omega_{2}=\omega_{f}$) the Hamiltonian in
the interaction picture is unchanged and is not time dependent
($\mbox{\boldmath$H^{I}$}=\mbox{\boldmath$\exp^{\frac{i}{\hbar}H_{0}t}H\exp^{-\frac{i}{\hbar}H_{0}t}$}=\mbox{\boldmath$H$};\
\mbox{\boldmath$H_{0}$}\equiv\mbox{\boldmath$H_{m}$}+\mbox{\boldmath$H_{f}$}$)
since $[\mbox{\boldmath$H_{0}$}, \mbox{\boldmath$V_{mf}$}]=0$ (when
there is no resonance one can still transform to an interaction
picture in which the Hamiltonian is unchanged \cite{Stenholm}).
However, as indicated previously (eq. \ref{eq:ixn_rep_h}), in the
interaction picture, the Hamiltonian part of the evolution of the
density matrix is only via the interaction term
$\mbox{\boldmath$V_{mf}$}=\mbox{\boldmath$H^{I}$}-
\mbox{\boldmath$H_{0}$}$.

Before we move on to discuss the ED-JCM as a quantum amplifier, we
wish to discuss the main differences between the ED-JCM and the
quantum theory of the laser due to Scully and Lamb (SL)
\cite{Scully00} \cite{Scully01}. Firstly, in the SL model the
material system (the atom) has either four levels \cite{Scully01} or
five levels \cite{S&Z}, whereas in the ED-JCM the matter has three
levels. Secondly, in the SL model the transitions between the two
upper lasing levels and the two lower levels is achieved through a
phenomenological decay, whereas in the ED-JCM population may also be
pumped from the ground state to the two upper lasing levels through
the full dissipative Lindblad super operator. Thirdly, in the SL
model the atom is assumed to be injected into the cavity in the
upper lasing level and interact with the cavity for a time $\tau$,
whereas in the ED-JCM the matter is in continuous contact with the
quantized cavity mode, and amplification is achieved for a wide
range of initial states. Finally, in the SL model the field is
allowed to decay using the Weiskopf-Wigner formalism, whereas in the
ED-JCM discussed in this paper the field does not decay. In
principle, cavity losses can be introduced to the ED-JCM. However,
we do not consider field damping in this paper, which allows us to
compare the thermodynamical fluxes in the quantum ED-JCM with their
analog in a semiclassical ED-JCM (section VII) and a similar model
by Geva and Kosloff in which field damping is not included
\cite{Geva03} \cite{Geva04}. The differences between our model and
that of the SL model will be seen below to play a crucial role in
our ability to give a thermodynamic foundation of amplification.

\section{The ED-JCM as a quantum amplifier}

The ED-JCM master equation, eq. \ref{MQE},  was solved using the
standard Runge-Kutta method (fourth-order \cite{Mathews}) for
various choices of parameters. The accuracy of the solution was
checked by decreasing the step size. Furthermore, in order to test
whether the numerical solution captures all time scales (especially
the rapid oscillations), the algorithm was tested on the simple JCM
\cite{JCM} which can be solved analytically \cite{Knight}
\cite{Erez01}. In all plots presented here
$\Gamma_{02}=\Gamma_{01}=\Gamma=0.001,\ \lambda=1,\ n_{02}=0.1,\
n_{01}=10$, and quantities are given in atomic units. The condition
$\lambda\gg\Gamma$ corresponds physically to a situation where the
coupling between the matter and the selected quantized cavity mode
is much stronger than the matter-reservoir coupling.

The energy flux of the full matter-field system and the individual
subsystems is given by:
\begin{eqnarray}
\dot{E}_{mf}&\equiv&\textrm{Tr}\{\mbox{\boldmath$\dot{\rho}_{mf}H$}\}=\textrm{Tr}\{\mathcal{L}_{d}[\mbox{\boldmath$\rho_{mf}$}]\mbox{\boldmath$H$}\}\nonumber\\
\dot{E}_{m}&\equiv&\textrm{Tr}\{\dot{\rho}_{m}H_{m}\}=-\frac{i}{\hbar}\textrm{Tr}\{\mbox{\boldmath$\rho_{mf}$}[\mbox{\boldmath$H_{m}$},
\mbox{\boldmath$V_{mf}$}]\}+\textrm{Tr}\{\mathcal{L}_{d}[\mbox{\boldmath$\rho_{mf}$}]\mbox{\boldmath$H_{m}$}\}=P_{m}+\dot{Q}_{m}\nonumber\\
\dot{E}_{f}&\equiv&\textrm{Tr}\{\dot{\rho}_{f}H_{f}\}=-\frac{i}{\hbar}\textrm{Tr}\{\mbox{\boldmath$\rho_{mf}$}[\mbox{\boldmath$H_{f}$},
\mbox{\boldmath$V_{mf}$}]\}=P_{f},\label{firstlawall}
\end{eqnarray}
where $\dot{E}_{mf}$, $\dot{E}_{m}$, and $\dot{E}_{f}$ are the
energy fluxes of the full matter-field system, the matter, and the
field, respectively. $\rho_{m}$ and $\rho_{f}$ are obtained from
$\mbox{\boldmath$\rho_{mf}$}$ by a partial trace over the field or a
partial trace over the matter, respectively. $H_m$ and $H_f$ are the
Hamiltonians of the matter and field subsystems, respectively,
without the tensor product with the identity. Note that the energy
fluxes of the individual subsystems in eq. \ref{firstlawall} are
defined via $\rho_{m}$ and $\rho_{f}$ together with the subsystem
Hamiltonians $H_{m}, H_{f}$. In the next two subsections we discuss
the transient and steady state energetic behavior of the ED-JCM.
Since there is direct dissipation only through matter-reservoir
coupling, there is no heat flux associated with the field (this is
physically expected, and was shown analytically elsewhere
\cite{Erez02}).

\subsection{Transient behavior}

\begin{figure}[htb]
\begin{center}
\includegraphics[width=8cm]{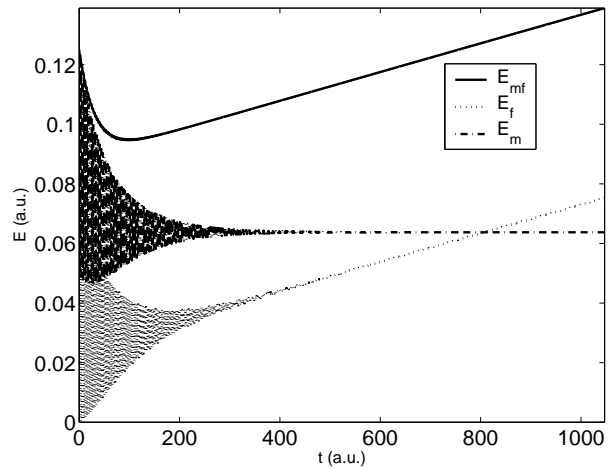}
\end{center}
\caption{Energy of the full matter-field system (solid line) and the
individual subsystem (field dotted line, matter dash-dot line) for
an initial state where the matter is in state $|1\rangle$ and the
selected cavity mode has no photons
($\mbox{\boldmath$\rho_{mf}(0)$}=(|1\rangle\langle
1|)_{m}\otimes(|0\rangle\langle 0|)_{f}$). Note that at long times
there is a steady state increase in the field's
energy.}\label{EDJCMenergy}
\end{figure}

The energy of the full matter-field system and of the individual
subsystems is plotted in Fig. \ref{EDJCMenergy} for an initial state
where the matter is in state $|1\rangle$ and the selected cavity
mode has no photons
($\mbox{\boldmath$\rho_{mf}(0)$}=(|1\rangle\langle
1|)_{m}\otimes(|0\rangle\langle 0|)_{f}$).
At short times, $t<\Gamma^{-1}_{eff}$, the matter and field energies
oscillate at a frequency of $\Omega=\pi\lambda$ \cite{Knight}. Here
$\Gamma_{eff}=\Gamma\frac{n_{01}+n_{02}}{2}$ is the effective decay
constant.

\begin{figure}[htb]
\begin{center}
\includegraphics[width=8cm]{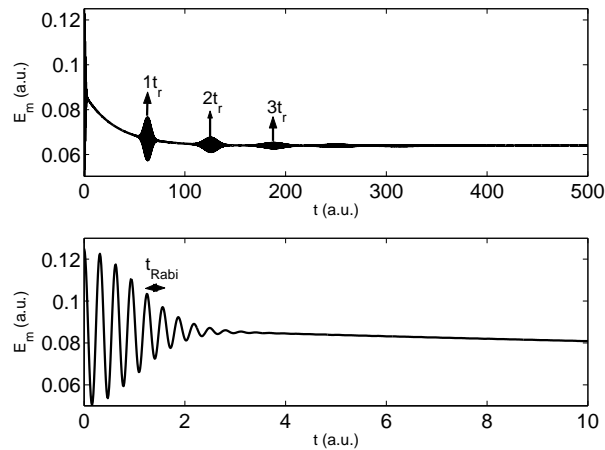}
\end{center}
\caption{Matter energy at short times ($t<(2\Gamma)^{-1}$). The
revival time (top) is $t_{r}=\frac{2\pi m\sqrt{\langle
n\rangle}}{\lambda}; \langle n\rangle=|\alpha|^{2}=100$, where $m$
is a positive integer. The Rabi oscillation time (bottom) is
$t_{{\rm Rabi}}=\frac{\pi}{\lambda\sqrt{\langle
n\rangle}}$.}\label{EDJCMenergyzoom}
\end{figure}

Moreover, at short times the well known collapse and revival
phenomena \cite{Eberly} \cite{Gea} is observed for a sufficiently
excited initial coherent state as depicted in Fig.
\ref{EDJCMenergyzoom}.

\begin{figure}[htb]
\begin{center}
\includegraphics[width=10cm]{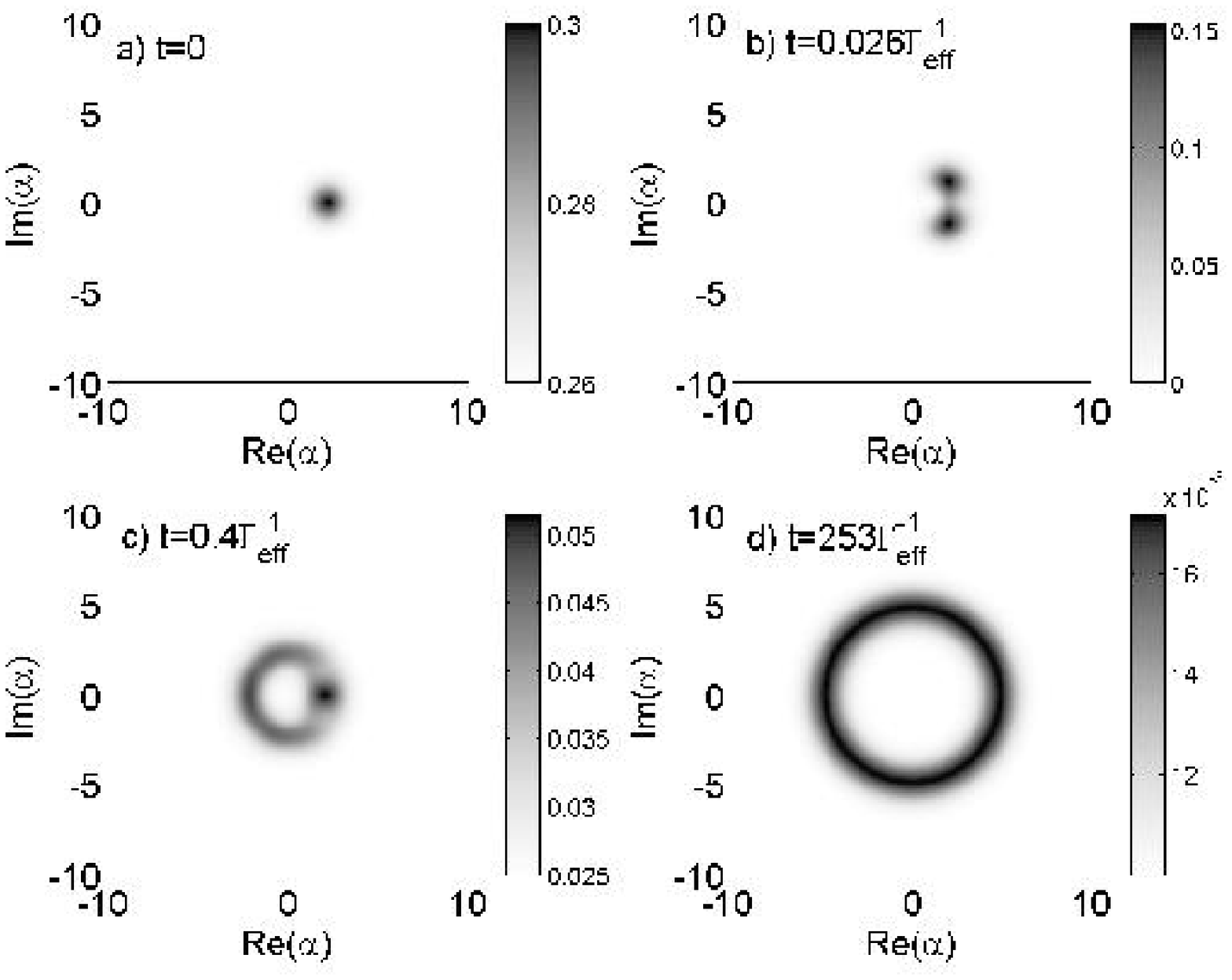}
\end{center}
\caption{Husimi-Kano $Q$ function of the selected cavity mode,
$\rho_{f}(0)=|\alpha\rangle\langle\alpha|; |\alpha|^{2}=5$. a): at
$t=0$ the $Q$ function is a narrow 2D gaussian. b) and c): at
$t=0.026\Gamma_{eff}^{-1},\ t=0.4\Gamma_{eff}^{-1}$ the $Q$ function
is spread in phase space inhomogeneously. d): at
$t=253\Gamma_{eff}^{-1}$ the $Q$ function has expanded (due to
amplification of the selected mode) into a radially symmetric
annulus (all the initial phase information is lost).}\label{EDJCMQ}
\end{figure}

In order to monitor the field's coherence we calculate the quantum
optical Husimi-Kano $Q$ function which is defined by
\cite{Schleich}:
\begin{equation}
Q(\alpha_{r},
\alpha_{i})=\frac{1}{\pi}\langle\alpha|\rho_{f}|\alpha\rangle,
\end{equation}
where $|\alpha\rangle$ is a (generally complex) coherent state. In
Fig. \ref{EDJCMQ} we plot the $Q$ function at four different times
($t=0$, $t=0.026\Gamma^{-1}_{eff}$, $t=0.4\Gamma_{eff}^{-1}$,
$t=253\Gamma^{-1}_{eff}$) for the initial state
$\mbox{\boldmath$\rho_{mf}(0)$}=(|1\rangle\langle
1|)_{m}\otimes(|\alpha \rangle \langle\alpha|)_f$, $|\alpha|^{2}=5$.
The $Q$ function at $t=0$ is that of a coherent state with the phase
centered around the real axis (Fig. \ref{EDJCMQ}a). At transient
times, the $Q$ function spreads in phase space, but it is not
homogeneous as seen in Fig. \ref{EDJCMQ}b and  Fig. \ref{EDJCMQ}c.

\subsection{Steady state behavior}

At much longer times, $t\geq\Gamma_{eff}^{-1}$, the matter energy
decreases to a steady state value, while the field energy increases
with a steady state power of $P_{f}^{ss}=4.5975e^{-5}$ (the
numerical value of a linear fit to the last $10000$ points,
$R^{2}=1.000$) as seen in Fig. \ref{EDJCMenergy}. Another indication
for an increase in the field's energy is seen in the steady state
increase in the full system energy. Thus, the field and the
matter-field system as a whole never reach a steady state for the
type of evolution discussed in this paper. The relation between the
full system energy and the energy of the individual component
subsystems will be discussed in the next sections. A steady state
increase in the field's energy is clearly an amplification of the
selected cavity mode. This behavior contrasts with the simple JCM in
which the atom and field oscillate forever (the atom oscillates
between the excited and ground states while the field oscillates
between the $|0\rangle$ and $|1\rangle$ Fock states). Amplification
of the selected cavity mode will occur with any other coherent
state, including the $|0\rangle$ Fock state. The fact that the
field's energy increases monotonically is not unreasonable, since
the harmonic oscillator is infinite, and since we do not consider
direct dissipation of the cavity mode (which could be modeled by
transmissive mirrors if desired).

The collapse and revival phenomenon at longer times  is completely
damped due to the dissipative contribution to the Liouvillian as
seen in Fig. \ref{EDJCMenergyzoom}.  At these long times all phase
(internal coherence) information is lost: the $Q$ function is
radially symmetric and is dispersed on a bigger area (bottom of Fig.
\ref{EDJCMQ}c). From this time onwards the shape of the $Q$ function
remains unchanged, and it expands fully symmetrically. The decay of
the initial field coherence is also reflected in the decay of the
off-diagonal field density matrix elements. At
$t=10\Gamma_{eff}^{-1}$, the off-diagonal matrix elements are
$10^{-13}$ times smaller than their initial value, and are
practically zero. All the remaining density matrix elements are
diagonal with a Poissonian-like photon distribution whose average
number of photons increases with time.

The full density matrix can be divided into a $3\times3$ block
matrix, each block associated with one element of the matter density
matrix.  At long times, $t\geq2\Gamma_{eff}^{-1}$, the matter-field
inter-coherence is maintained by the non-vanishing matrix elements:
$\rho_{1;n,2;n+1}, \rho_{2;n+1,1;n}$ of the full density matrix.
These elements correspond to matter-field coupling, maintained via
the structure of the JCM Hamiltonian.  Other density matrix elements
at these long times are $5-9$ orders of magnitude smaller than the
dominant non vanishing elements discussed above.

\section{Entropy in the ED-JCM}

We consider now the entropy in the ED-JCM. In the next two
subsections we discuss the transient and steady state entropic
behavior. In Subsection C we discuss the relation between the
entropies of the individual subsystems and entanglement, both at
transient and steady state times.

\subsection{Transient behavior}

\begin{figure}[htb]
\begin{center}
\includegraphics[width=10cm]{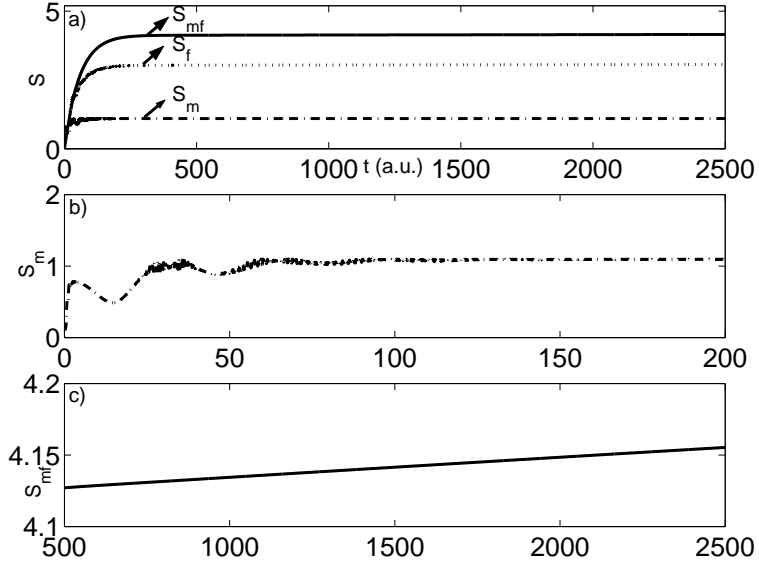}
\end{center}
\caption{Entropy of the ED-JCM for the initial state
$\rho_{mf}(0)=(|1\rangle\langle1|)_{m}\otimes(|\alpha\rangle\langle\alpha|;|\alpha|^{2}=25)_{f}$.
a): entropy of the full matter-field system (solid line) and the
individual subsystem (field dotted line, matter dash-dot line) for
the full evolution. b): matter entropy for times
$t\leq\frac{\Gamma_{eff}^{-1}}{2}$. c): entropy of the full
matter-field system for times
$t\geq2\Gamma_{eff}^{-1}$.}\label{EDJCMentropy}
\end{figure}

In Fig. \ref{EDJCMentropy} we plot the entropy of the full
matter-field system and the individual subsystems for the initial
state
$\mbox{\boldmath$\rho_{mf}(0)$}=(|1\rangle\langle1|)_{m}\otimes(|\alpha\rangle\langle\alpha|)_f$,
$|\alpha|^{2}=25$.
The entropy plots at the top of Fig. \ref{EDJCMentropy} show that
there is a rapid rise ($\tau\approx\frac{\Gamma_{eff}^{-1}}{2}$) in
the entropies of the matter (dash-dot line), the field (dotted
line), and the full matter-field system (solid line). This overall
rise in entropy is discussed in Subsection C) and is attributed to
the dissipative nature of the problem.

At times $t<\Gamma_{eff}^{-1}$, the entropy of the individual
subsystems (matter or field) is oscillatory, as seen by the matter
entropy plot in the middle of Fig. \ref{EDJCMentropy}. This behavior
is typical of the simple JCM \cite{Knight} \cite{Erez01}. At this
stage $S_{m}+S_{f}>S_{mf}$; in Subsection C) we attribute the excess
entropy to entanglement.

\subsection{Steady state}

Fig. \ref{EDJCMentropy}a suggests that at $t>\Gamma_{eff}^{-1}$ the
matter-field system has reached a steady state. Indeed, at times
$t>\Gamma_{eff}^{-1}$ the energy of the matter remains constant
(dash-dot line in Fig. \ref{EDJCMenergy}). However, as was indicated
in the previous section, the field energy plot (dotted line in Fig.
\ref{EDJCMenergy}) and the matter-field energy plot (solid line in
Fig. \ref{EDJCMenergy}) both show a constant rise for
$t\geq\Gamma_{eff}^{-1}$. Furthermore, a closer inspection of the
matter-field entropy (Fig. \ref{EDJCMentropy}b) reveals a constant
slight rise in entropy at times $t\geq\Gamma_{eff}^{-1}$ (a similar
rise in the field entropy is also observed). Moreover, the field
density matrix eigenvalues change in the second and third
significant figures over a $\Gamma_{eff}^{-1}$ time scale. These
findings give further proof of the fact that the field and
matter-field system as a whole never reach a steady state.

\subsection{Entanglement}

We will now analyze the nature of the entropies associated with the
subsystems.  The entropy of the individual parts of a bipartite
system is closely tied to the issue of entanglement \cite{Cerf}
\cite{Erez01}. An important measure for entanglement is the
conditional entropy, defined for the matter-field system by:
\begin{eqnarray}
S(m|f)&\equiv&S_{mf}-S_{f}\nonumber\\
S(f|m)&\equiv&S_{mf}-S_{m},
\end{eqnarray}
where $S(m|f)$ is the conditional entropy of the matter, and
$S(f|m)$ is the conditional entropy of the field.  In contrast with
the conditional entropy in classical bipartite systems, the
conditional entropy in quantum bipartite systems can assume negative
values. In this case, the correlation between the two parts of the
system is of a purely quantum nature, and the system is therefore
entangled. A fine example for entanglement in the context of our
work is the simple JCM. Consider an initial state given by:
$\mbox{\boldmath$\rho_{mf}(0)$}=(|e\rangle\langle
e|)_{a}\otimes(|0\rangle\langle 0|)_{f}$, where the atom (indicated
by subscript $a$) is in the excited state and the cavity mode is
empty. Under the JCM Hamiltonian (pure Hamiltonian dynamics), the
full atomic-field entropy is constant $S_{af}(t)=0$. However, during
most of the evolution time the conditional entropies of both the
atom and field (which are equal) are negative. In our case, at short
times, $0<t\leq\frac{\Gamma^{-1}_{eff}}{5}$ we find that the
matter's conditional entropy is negative. Thus, the excess entropy
$S_{m}+S_{f}>S_{mf}$ at these times is attributed to entanglement.

A more powerful test for entanglement, introduced originally by
Peres \cite{Peres}, is the negativity of the partially transposed
density matrix. The partially transposed density matrix is defined
by:
\begin{equation}
\mbox{\boldmath$\rho_{i\alpha,j\beta}^{T_{2}}$}\equiv
\mbox{\boldmath$\rho_{i\beta,j\alpha}$}.
\end{equation}
A sufficient condition for entanglement is the negativity of
$\mbox{\boldmath$\rho^{T_{2}}$}$. However, since this test applies
only to finite dimensional density matrices, one should take care
not to mistake truly negative eigenvalues with negative eigenvalues
that are an artifact of truncation of an infinite Hilbert space
\cite{Erez01}. Indeed, at times smaller than the typical decay time
($\tau=\Gamma_{eff}^{-1}$), we find that the matter-field partially
transposed density matrix $\mbox{\boldmath$\rho_{mf}^{T_{2}}$}$ is
negative (negative eigenvalues with a substantial absolute value are
found up to a time $t\approx\frac{3\Gamma_{eff}^{-1}}{4}$, and hence
the matter-field system is entangled.

At $t>\Gamma_{eff}^{-1}$ the matter and field conditional entropies
are positive. Moreover, the conditional entropies are almost equal
to the partial entropies:
$\frac{S(m|f)}{S_{m}},\frac{S(f|m)}{S_{f}}>99\%$, and
$\mbox{\boldmath$\rho_{mf}^{T_{2}}$}$ is positive (as was indicated
before). All these findings lead us to conclude that in all
likelihood at long times the matter-field system is only weakly
classically correlated.

We summarize this section by stating that at short times
($t\leq\Gamma_{eff}^{-1}$), when the partial entropies are
oscillatory (see bottom of Fig. \ref{EDJCMentropy}), the
matter-field system is entangled, as verified by the negative
conditional entropies and the negative partially transposed full
density matrix. However, as dissipation sets in, the matter-field
system becomes less and less entangled. At $t>\Gamma_{eff}^{-1}$,
when the partial entropies are not oscillating any more, the
matter-field system in all likelihood is not entangled (as verified
by the positive conditional entropies and the positive partially
transposed full density matrix), and the overall rise in entropy is
attributed to the dissipative Lindblad super operator.

\section{Thermodynamic analysis of the steady state solution}

\subsection{The first law}

The first law of thermodynamics is essentially given in equation
\ref{firstlawall}. However, some fine details need more
clarification. The first law of thermodynamics for the full
matter-field system in differential form is given by:
\begin{equation}
\dot{E}_{mf}\equiv\textrm{Tr}\{\mbox{\boldmath$\dot{\rho}_{mf}H$}\}=\textrm{Tr}\{\mathcal{L}_{d}[\mbox{\boldmath$\rho_{mf}$}]H\}=\dot{Q}_{m}+\dot{Q}_{f}+\dot{Q}_{V}=\dot{Q}_{m}+\dot{Q}_{V},
\end{equation}
where
$\dot{Q}_{f}\equiv\textrm{Tr}\{\mathcal{L}_{d}[\mbox{\boldmath$\rho_{mf}$}]\mbox{\boldmath$H_{f}$}\}=0$
as was shown elsewhere \cite{Erez02}, and
$\dot{Q}_{V}\equiv\textrm{Tr}\{\mathcal{L}_{d}[\mbox{\boldmath$\rho_{mf}$}]\mbox{\boldmath$V_{mf}$}\}$.
$\dot{Q}_{m}\equiv\textrm{Tr}\{\mathcal{L}_{d}[\mbox{\boldmath$\rho_{mf}$}]\mbox{\boldmath$H_{m}$}\}=\textrm{Tr}\{\mathcal{L}_{dC}[\mbox{\boldmath$\rho_{mf}$}]\mbox{\boldmath$H_{m}$}\}+\textrm{Tr}\{\mathcal{L}_{dH}[\mbox{\boldmath$\rho_{mf}$}]\mbox{\boldmath$H_{m}$}\}=\dot{Q}_{mC}+\dot{Q}_{mH}$
is the heat flux associated with the matter and it is composed of
heat fluxes from/to the cold and hot heat reservoirs. Note that to
an observer looking on the matter-field system as a whole, the full
system is only dissipating heat.

Another way to formulate the first law of thermodynamics is based on
the energy flux of individual subsystems. The first law of
thermodynamics for the matter and field separately (in differential
form) is given by:
\begin{eqnarray}
\dot{E}_{m}&\equiv&\textrm{Tr}\{\dot{\rho}_{m}H_{m}\}=-\frac{i}{\hbar}\textrm{Tr}\{\mbox{\boldmath$\rho_{mf}$}[\mbox{\boldmath$H_{m}$},
\mbox{\boldmath$V_{mf}$}]\}+\textrm{Tr}\{\mathcal{L}_{d}[\mbox{\boldmath$\rho_{mf}$}]\mbox{\boldmath$H_{m}$}\}=P_{m}+\dot{Q}_{m}\label{emdot}\\
\dot{E}_{f}&\equiv&\textrm{Tr}\{\dot{\rho}_{f}H_{f}\}=-\frac{i}{\hbar}\textrm{Tr}\{\mbox{\boldmath$\rho_{mf}$}[\mbox{\boldmath$H_{f}$},
\mbox{\boldmath$V_{mf}$}]\}=P_{f},\label{efdot}
\end{eqnarray}
where
$P_{m}\equiv-\frac{i}{\hbar}\textrm{Tr}\{\mbox{\boldmath$\rho_{mf}$}[\mbox{\boldmath$H_{m}$},
\mbox{\boldmath$V_{mf}$}]\}$, and
$P_{f}\equiv-\frac{i}{\hbar}\textrm{Tr}\{\mbox{\boldmath$\rho_{mf}$}[\mbox{\boldmath$H_{f}$},
\mbox{\boldmath$V_{mf}$}]\}$ are the power terms. Since we are
considering the case of perfect matter-field resonance,
$P_{m}=-P_{f}$
($[\mbox{\boldmath$H_{m}$},\mbox{\boldmath$V_{mf}$}]=-[\mbox{\boldmath$H_{f}$},\mbox{\boldmath$V_{mf}$}]$),
hence:
\begin{equation}
\dot{E}_{m}+\dot{E}_{f}=\dot{E}_{mf}-\dot{Q}_{V}.\label{fluxsum}
\end{equation}
It may be shown that $\dot{Q}_{V}$ vanishes if the off-diagonal
matrix elements of $\mbox{\boldmath$\rho_{mf}$}$ are purely
imaginary. Note that to an observer looking on the matter alone work
flux (power) and heat fluxes are identified according to eq.
\ref{emdot}, in agreement with the traditional thermodynamic
partitioning of energy into work and heat. The field, which is the
work source, either receives or emits energy to the working medium
(the matter) in the form of power.  In this paper we are interested
in optical amplification.  Under such conditions, at steady state
the energy flux balance is such that
$P_{m}^{ss}<0,\dot{Q}_{mH}>0,\dot{Q}_{mC}<0$ and the three-level
system operates thermodynamically as a heat engine.

\subsection{Second law. Clausius formulation}

The second law of thermodynamics is obtained via the entropy
production function of the full bipartite matter-field system, which
is defined by \cite{Spohn}, \cite{Alicki}:
\begin{equation}
\sigma\equiv\frac{\partial S_{mf}}{\partial t}+J,\label{entprod0}
\end{equation}
where $\frac{\partial S_{mf}}{\partial t}$ is the entropy production
associated with the bipartite matter-field density matrix (via
differentiation of the von Neumann entropy), and $J$ is the entropy
production associated with the reservoirs (via the heat flux from/to
the reservoirs) given by:
\begin{equation}
J=-\beta_{C}\dot{Q}_{C}-\beta_{H}\dot{Q}_{H},
\end{equation}
where $\beta_{C(H)}=(k_{B}T_{C(H)})^{-1}$, and
$\dot{Q}_{C(H)}\equiv\textrm{Tr}\{\mathcal{L}_{dC(H)}[\mbox{\boldmath$\rho_{mf}$}](\mbox{\boldmath$H_{m}$}+\mbox{\boldmath$V_{mf}$})\}=\dot{Q}_{mC(H)}+\dot{Q}_{VC(H)}$.
Spohn showed that for a completely positive map (such as the
Lindblad super operator) \cite{Spohn}:
\begin{equation}
\sigma^{Q}\geq0.\label{entprod1}
\end{equation}
Equation \ref{entprod1} represents the differential form of the
second law of thermodynamics in Clausius's formulation, since the
sum of the entropy changes of the system and reservoirs is
guaranteed to be positive.

\subsection{Second law. Carnot's formulation}

We now define a new entropy production function:
\begin{equation}
\sigma_{m}\equiv\frac{\partial S_{m}}{\partial
t}+J_{m},\label{entprodmat}
\end{equation}
where $\frac{\partial S_{m}}{\partial t}$ is the entropy production
associated with the matter density matrix (via differentiation of
the matter von Neumann entropy), and $J_{m}$ is the entropy
production associated with the reservoirs,
\begin{equation}
J_{m}=-\beta_{C}\dot{Q}_{mC}-\beta_{H}\dot{Q}_{mH},\label{Jm}
\end{equation}
taking into account the contribution only from the matter heat flux
\begin{equation}
\dot{Q}_{m}=\dot{Q}_{mC}+\dot{Q}_{mH}.\label{dotQm}
\end{equation}
The physical idea behind $\sigma_{m}$ is that it is built only from
matter thermodynamic fluxes: the intrinsic entropy flux
$\frac{\partial S_{m}}{\partial t}$ and the entropy flux $J_{m}$
arising just from matter heat fluxes. For many initial matter states
$\dot{Q}_{VC(H)}=0$ at all times, and hence $J_{m}=J$. Moreover,
when the matter reaches a steady state we always find numerically
(irrespective of the initial matter state) that $\dot{Q}_{VC(H)}=0$.
This is the case also in the semiclassical ED-JCM discussed in
section VII, where it can be shown analytically that
$\dot{Q}_{VC(H)}^{ss}=0$. Therefore, at steady state, $\sigma_{m}$
is physically similar to the entropy production function in the
semiclassical case, $\sigma^{SC}$, where the field is not quantized.
In contrast with $\sigma^{SC}$ and $\sigma^{Q}$, $\sigma_{m}$ is not
guaranteed to be positive at all times (especially at times
$t<(2\Gamma)^{-1}$, due to the highly oscillatory nature of the
partial entropy at short times). However, when the matter reaches a
steady state ($\frac{\partial S_{m}}{\partial t}=0$), the increase
in the field's entropy is marginal (as was indicated before), and
the main source of entropy production is the heat flux from/to the
heat reservoirs ($J>\frac{\partial S_{mf}}{\partial t}$). Thus, when
the matter reaches a steady state
\begin{equation}
\sigma_{m}=J_{m}=J>0,\label{sigmss}
\end{equation}
and since the matter operates in a heat engine mode
($\dot{Q}_{mH}>0$ and $\dot{Q}_{mC}<0$), we obtain Carnot's
efficiency formula:
\begin{equation}
\eta\equiv-\frac{P_{a}}{\dot{Q}_{mH}}=\frac{\dot{Q}_{mC}+\dot{Q}_{mH}}{\dot{Q}_{mH}}\leq\frac{T_{H}-T_{C}}{T_{H}},\label{Carnot}
\end{equation}
where we have used eq. \ref{emdot} with eq. \ref{dotQm}, and eq.
\ref{sigmss} with eq. \ref{Jm}. For example, the efficiency of the
heat engine for the choice of parameters discussed in the previous
plots and for various initial field strengths (ranging from an empty
cavity up to $100$ photons) is $75\%$, which is less than the Carnot
efficiency which is $99\%$.

Scovil and Schulz-DuBois gave an intuitive, but non-thermodynamics
definition of the efficiency of the three-level system operating as
a maser \cite{Scovil01}:
\begin{equation}
\eta_{M}=\frac{\omega_{s}}{\omega_{p}}, \label{Scovil_djt}
\end{equation}
where $\omega_{s}$ is the signal (maser) frequency, and $\omega_{p}$
is the pump frequency (central frequency of the hot reservoir,
$\omega_1 - \omega_0$ ). By substituting our initial choice of
parameters ($\omega_{s}=\omega=0.075$, and
$\omega_{p}=\omega_{1}-\omega_{0}=0.1$) we see that our numerical
result agrees precisely with the efficiency estimated by Scovil and
Schulz-DuBois. We find numerically that the efficiency calculated
from eq. \ref{Carnot} agrees precisely with the efficiency
calculated using eq. \ref{Scovil_djt}. Although we do not have
analytic proof of this equivalence for the case of quantized light,
we show below that this equivalence can be derived analytically when
the light is treated classically.

\subsection{The ED-JCM: A work source with an entropy content}

A work source is the physical entity on which work is done, or which
performs work on a system (working medium). Conventional wisdom in
classical thermodynamics states that a work source's entropy is
constant during the operation of a heat engine \cite{Reif}.  Whether
the classical engine operates cyclically, as in the usual Carnot
cycle, or synchronously, the working medium returns to its initial
state.  The working assumption in thermodynamics is that entropy may
be produced at the boundary of the working medium and the heat
reservoirs, but not at the boundary with the work source.

The work source in the quantum amplifier discussed in this paper is
the selected cavity mode which is amplified. At steady state, the
density matrix of the matter becomes constant and thus its entropy
is unchanged from this time onwards.  Since energy is flowing from
the hot reservoir to the cold reservoir and work is produced in the
form of amplification of the cavity mode, this corresponds to the
engine operating in synchronous mode with the cavity mode as the
work source. Inspection of Fig. \ref{EDJCMentropy} shows that the
entropy of the light is not constant.  Even after the matter reaches
a steady state, the entropy of the light continues to grow linearly
in time.

\section{The semiclassical ED-JCM}

\subsection{Equations of motion}

The semiclassical ED-JCM master equation is similar to the quantum
ED-JCM master equation given by equation \ref{MQE}. However, since
the selected quantized cavity mode is replaced by a time dependent
field, major differences arise. The field is considered as an
external degree of freedom, and hence it has no entropy content. We
propagate a $3\times3$ density matrix representing the matter only,
and all operators are represented by $3\times3$ matrices (as opposed
to $(3\otimes n)\times(3\otimes n)$ in the fully quantized case).
Finally, the Hamiltonian part of the Liouvillian assumes a different
form, where the creation and annihilation field operators are
replaced by clockwise and anti-clockwise oscillating exponents.
Despite the last difference, we note that in perfect matter-field
resonance the Hamiltonian part of the Liouvillian in the interaction
picture is time independent. The semiclassical Hamiltonian is given
by:
\begin{equation}
H=H_{m}+V,\label{EDJCMsc}
\end{equation}
where $H_{m}$ is the matter Hamiltonian as given in equation
\ref{JCMH} (without the tensor product with $\openone_{f}$), and
\begin{equation}
V=\lambda_{sc}(\sigma_{21}\rm{e}^{i\omega
t}+\sigma_{21}^{\dag}\rm{e}^{-i\omega t})
\end{equation}
is the interaction Hamiltonian with a classical single coherent mode
in the RWA. $\lambda_{sc}$ is the semiclassical matter-field
coupling constant (which can be obtained via the semiclassical
coupling matrix element \cite{Loudon}) given by (atomic units):
\begin{equation}
\lambda_{sc}=\mathbf{\hat{D}}\cdot\mathbf{\hat{\epsilon}}\frac{E_{0}}{2},\label{lsc}
\end{equation}
where $\mathbf{\hat{D}}$ is the dipole operator,
$\mathbf{\hat{\epsilon}}$ is the field polarization, and $E_{0}$ is
the field amplitude which can be estimated by calculating the
average value of the quantum field operator for a coherent state
\cite{Loudon}:
\begin{equation}
E_{0}=\left(\frac{8\pi\omega}{\bar{V}}\right)^{-1/2}|\alpha|,\label{E0}
\end{equation}
where $\omega$ is the mode frequency (not necessarily in resonance
with the atomic transition), $\bar{V}$ is the cavity volume, and
$|\alpha|$ is the field strength. The quantum matter-field coupling
constant is given by (atomic units) \cite{Loudon}:
\begin{equation}
\lambda=\mathbf{\hat{D}}\cdot\mathbf{\hat{\epsilon}}\left(\frac{2\pi\omega}{\bar{V}}\right)^{-1/2}.\label{lq}
\end{equation}
Combining equations \ref{lsc}, \ref{E0}, and \ref{lq} we obtain
that:
\begin{equation}
\lambda_{sc}=\lambda|\alpha|.
\end{equation}
The dissipative part of the Liouvillian is identical to equation
\ref{Ldch} (without the tensor product with $\openone_{f}$).

Substitution of equation \ref{EDJCMsc} and equations \ref{Ldch} (not
in tensor product form) into equation \ref{MQE} yields a set of
equations for the matter density matrix elements. In the interaction
picture (with $H_0$ given by $H_m$) and assuming perfect
matter-field resonance, these equations take the form:
\begin{eqnarray}
\dot{\rho}_{00}&=&2\Gamma_{01}(n_{01}+1)\rho_{11}-2\Gamma_{01}n_{01}\rho_{00}-2\Gamma_{02}n_{02}\rho_{00}+2\Gamma_{02}(n_{02}+1)\rho_{22}\nonumber\\
\dot{\rho}_{11}&=&-i\lambda_{sc}\rho_{21}+i\lambda_{sc}\rho_{12}-2\Gamma_{01}(n_{01}+1)\rho_{11}+2\Gamma_{01}n_{01}\rho_{00}\nonumber\\
\dot{\rho}_{22}&=&-i\lambda_{sc}\rho_{12}+i\lambda_{sc}\rho_{21}-2\Gamma_{02}(n_{02}+1)\rho_{22}+2\Gamma_{02}n_{02}\rho_{00}\nonumber\\
\dot{\rho}_{12}&=&-i\lambda_{sc}\rho_{22}+i\lambda_{sc}\rho_{11}-\Gamma_{01}(n_{01}+1)\rho_{12}-\Gamma_{02}(n_{02}+1)\rho_{12}\nonumber\\
\dot{\rho}_{01}&=&i\lambda_{sc}\rho_{02}-\Gamma_{01}(2n_{01}+1)\rho_{01}-\Gamma_{02}n_{02}\rho_{01}\nonumber\\
\dot{\rho}_{02}&=&i\lambda_{sc}\rho_{01}-\Gamma_{02}(2n_{02}+1)\rho_{02}-\Gamma_{01}n_{01}\rho_{02}\nonumber\\
\dot{\rho}_{21}&=&\dot{\rho}_{12}^{*}\nonumber\\
\dot{\rho}_{10}&=&\dot{\rho}_{10}^{*}\nonumber\\
\dot{\rho}_{20}&=&\dot{\rho}_{20}^{*}.
\end{eqnarray}

\subsection{Thermodynamics of unipartite systems}

Heat flux ($\dot{Q}$) and power ($P$) for unipartite systems with
external (time dependent) forcing were originally defined by Alicki
\cite{Alicki}:
\begin{eqnarray}
\dot{Q}&=&\textrm{Tr}\left\{\frac{\partial\rho}{\partial
t}H\right\}=\textrm{Tr}\{\mathcal{L}_{d}[\rho]H\}\label{QAl}\\
P&=&\textrm{Tr}\left\{\rho\frac{\partial H}{\partial
t}\right\}\label{PAl}.
\end{eqnarray}

\subsection{Steady state solution of the semiclassical ED-JCM}

Before we derive the steady state power and heat flux we wish to
discuss the main differences between the semiclassical ED-JCM and
the semiclassical theory of the laser due to Lamb
\cite{Lambsclaser}. Firstly, in Lamb's model the material system
(the atom) has two levels, whereas in the semiclassical ED-JCM the
matter has three levels. Secondly, in Lamb's model pumping and decay
of the two lasing levels are phenomenological (where the pumping
function affects the field and thus the interaction term in the
Hamiltonian), whereas in the semiclassical ED-JCM pumping and
dumping of matter population from the ground state to the two upper
lasing levels is achieved through the full dissipative Lindblad
superoperator. Thirdly, in Lamb's model the field is allowed to
decay phenomenologically, whereas in the semiclassical ED-JCM
discussed in this paper the field does not decay. Finally, in Lamb's
model, Maxwell's equations for the classical field are solved self-
consistently with a quantum perturbative solution of the atomic
density matrix, whereas in the semiclassical ED-JCM discussed here
the field is not accounted for directly. As was mentioned
previously, in the semiclassical model of Geva and Kosloff the field
is not accounted for directly as well. Therefore, cavity damping is
not incorporated, and negative steady state power in the atom
signifies an increase in the field's energy.

The steady state solutions for $\rho_{01}$ and $\rho_{02}$ is
$\rho_{01}=\rho_{02}=0$, since $\gamma\rho_{01}=0$,
$\rho_{02}=\beta\rho_{01}$, and $\beta,\gamma>0$ (after applying the
steady state condition $\dot{\rho}_{02}=\dot{\rho}_{01}=0$).
Combining the equations for $\dot{\rho}_{12}$ and $\dot{\rho}_{21}$
at steady state ($\dot{\rho}_{12}=\dot{\rho}_{21}=0$) yields a
central equation:
\begin{equation}
|\rho_{12}|\cos\phi(\Gamma_{01}(n_{01}+1)+\Gamma_{02}(n_{02}+1))=0,
\end{equation}
where $\phi$ is the phase of the $\rho_{12}$ density matrix element.
There are now three possible physical solutions.

A. $|\rho_{12}|=0$. This yields:
\begin{eqnarray}
\rho_{00}&=&\frac{1}{1+2z}\nonumber\\
\rho_{11}&=&\rho_{22}=z\rho_{00},
\end{eqnarray}
where $z=\frac{n_{01}}{n_{01}+1}=\frac{n_{02}}{n_{02}+1}$. Note that
this corresponds to a very specific choice of parameters.

B. $\rm{cos}\phi=0,\ \phi=\frac{3\pi}{2}$. This yields a situation
where there is no inversion of the atomic levels:
\begin{equation}
\rho_{11}-\rho_{22}=-\frac{|\rho_{12}|}{\lambda_{sc}}(\Gamma_{01}(n_{01}+1)+\Gamma_{02}(n_{02}+1))<0.
\end{equation}
Moreover, it leads to a positive atomic steady state power which
corresponds to attenuation of the electromagnetic field. This is
outside the scope of the current paper, and will be explored in more
detail elsewhere \cite{Erez04}.

C. $\rm{cos}\phi=0,\ \phi=\frac{\pi}{2}$. This yields a situation
where there is an inversion of the atomic levels:
\begin{equation}
\rho_{11}-\rho_{22}=\frac{|\rho_{12}|}{\lambda_{sc}}(\Gamma_{01}(n_{01}+1)+\Gamma_{02}(n_{02}+1))>0.\label{deltapi2}
\end{equation}
The steady state solutions for the $\rho_{11}$, $\rho_{22}$, and
$|\rho_{12}|$ density matrix elements is obtained through the
solution of the following set of equations:
\begin{equation}
\left(
\begin{array}{ccc}
  -\Gamma_{01}(2n_{01}+1) & -\Gamma_{01}n_{01} & -\lambda_{sc} \\
  -\Gamma_{02}n_{02} & -\Gamma_{02}(2n_{02}+1) & \lambda_{sc} \\
  \lambda_{sc} & -\lambda_{sc} & -\frac{\Gamma_{01}(n_{01}+1)+\Gamma_{02}(n_{02}+1)}{\lambda_{sc}} \\
\end{array}
\right)\left(
\begin{array}{c}
         \rho_{11} \\
         \rho_{22} \\
         |\rho_{12}| \\
       \end{array}
\right)=\left(
\begin{array}{c}
  -\Gamma_{01}n_{01} \\
  -\Gamma_{02}n_{02} \\
  0 \\
\end{array}
\right)\label{atomicssmat}.
\end{equation}
The solution of equation \ref{atomicssmat} can be written as:
\begin{equation}
\rho^{ss}=\left(
\begin{array}{ccc}
  \rho_{00} & 0 & 0 \\
  0 & \rho_{11} & i|\rho_{12}| \\
  0 & -i|\rho_{12}| & \rho_{22} \\
\end{array}
\right)=\left(
\begin{array}{ccc}
  A/F & 0 & 0 \\
  0 & B/F & iD/F \\
  0 & -iD/F & C/F  \\
\end{array}
\right),\label{analytic}
\end{equation}
where $A,B,C,D,F$ are given in the Appendix II.

\subsection{Steady state heat flux and power in the semiclassical
ED-JCM}

We are now in position to compare the steady state heat fluxes and
power of the fully quantum model with the analytical solutions of
the semiclassical model. Applying Alicki's definitions (eq.
\ref{QAl} and eq. \ref{PAl}) to the semiclassical ED-JCM at steady
state yields:
\begin{eqnarray}
P^{ss}&=&-\frac{2\Gamma_{01}\Gamma_{02}\lambda_{sc}^3(n_{01}-n_{02})\omega}{F}\nonumber\\
\dot{Q}_{H}^{ss}&=&\frac{2\Gamma_{01}\Gamma_{02}\lambda_{sc}^3(n_{01}-n_{02})(\omega_{1}-\omega_{0})}{F}\nonumber\\
\dot{Q}_{C}^{ss}&=&-\frac{2\Gamma_{01}\Gamma_{02}\lambda_{sc}^3(n_{01}-n_{02})(\omega_{2}-\omega_{0})}{F},
\end{eqnarray}
where $F=F(\Gamma_{01},\Gamma_{02},n_{01},n_{02},\lambda_{sc})$ is a
positive constant given in the appendix. We note that at steady
state $\textrm{Tr}\{\mathcal{L}_{dH(C)}[\rho^{ss}]V\}=0$, and thus
$\dot{Q}_{H(C)}^{ss}=\textrm{Tr}\{\mathcal{L}_{dH(C)}[\rho^{ss}]H_{m}\}$.

Under the condition $\lambda\gg\Gamma$, the reservoir heat fluxes
and power for the fully quantum ED-JCM are found numerically to be
independent of $|\alpha|$ for the range $0\leq|\alpha|\leq10$ (which
corresponds to an initial coherent state ranging from no photons at
all to $100$ photons in the cavity). There are $0.5\%$ deviations
for the higher field strength range (where the initial number of
photons in the cavity is close to $100$) due to a slightly rougher
truncation of the Fock space.

The analytical semiclassical hot reservoir heat flux and power in
the range $0.1\leq|\alpha|\leq10$ are practically independent of
$|\alpha|$, and agree almost perfectly with the numerical steady
state fluxes in the fully quantum model. However, as $|\alpha|$
decreases below $0.1$ the semiclassical reservoir heat fluxes and
power change dramatically. This is of course expected, as
$\lambda_{sc}\propto|\alpha|$, and thus when the field's amplitude
decreases below $0.1$, $\lambda_{sc}$ is no longer much bigger than
$\Gamma$. Under the condition $\lambda\gg\Gamma$, we find
essentially perfect agreement between the numerical steady state
fluxes of the fully quantum ED-JCM and the analytical steady state
fluxes of the semiclassical ED-JCM. Therefore we can state that as
far as thermodynamical fluxes are considered, the semiclassical
ED-JCM captures the true physical picture. One important exception
is that in the semiclassical treatment, if there is no initial field
present at all ($E_{0}=0$), amplification can not take place.

\begin{figure}[htb]
\begin{center}
\includegraphics[width=10cm]{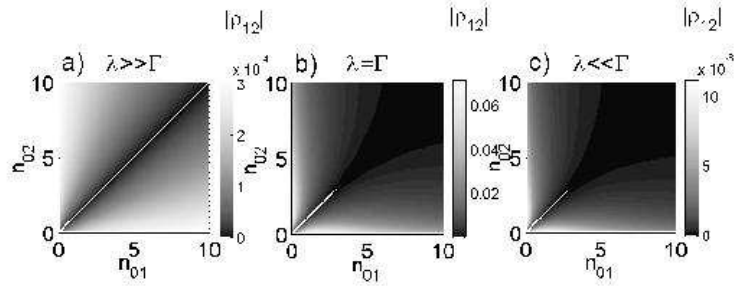}
\end{center}
\caption{Semiclassical atomic coherence. a) $\lambda\gg\Gamma$. b)
$\lambda=\Gamma$. c) $\lambda\ll\Gamma$.}\label{sccoherence}
\end{figure}

For completeness, we note that the steady state amplification
described above is only one of several thermodynamic modes of
operation of the light-matter system. Consider the expression for
the steady state power:
\begin{equation}
P^{ss}=\textrm{Tr}\left\{\rho^{ss}\frac{\partial V}{\partial
t}\right\}=-2\lambda_{sc}\omega|\rho^{ss}_{12}|,\label{powerpi2}
\end{equation}
where
$|\rho^{ss}_{12}|=\frac{\lambda_{sc}^{2}\Gamma_{01}\Gamma_{02}(n_{01}-n_{02})}{E}$.
A mathematically feasible solution for $|\rho^{ss}_{12}|$ is
obtained only when $n_{01}>n_{02}$ ($E$ is a positive constant).
Substituting $|\rho^{ss}_{12}|$ into equations \ref{deltapi2} and
\ref{powerpi2} reveals that atomic inversion and amplification go
together hand in hand. Inversion in the two excited state levels
implies negative power (corresponding to an amplification of the
electromagnetic field) and vice versa.  However, a full solution of
case B reveals that
$|\rho^{ss}_{12}|=-\frac{\lambda_{sc}^{2}\Gamma_{01}\Gamma_{02}(n_{01}-n_{02})}{E}$.
In this case, a mathematically feasible solution for
$|\rho^{ss}_{12}|$ is obtained only when $n_{02}>n_{01}$. Therefore,
$|\rho^{ss}_{12}|$ is a symmetric function of $|n_{01}-n_{02}|$. The
absolute value of the atomic coherence at steady state is plotted in
Fig. \ref{sccoherence} for three parameter ranges.  The different
thermodynamic modes of operation will be described in more detail in
a forthcoming publication \cite{Erez04}.
It can be seen that substantial atomic coherence is observed only
when $\lambda\approx\Gamma$.

\subsection{Engine efficiency}

What about the engine's efficiency? In the previous section we
mentioned that the (numerical) efficiency of the quantum amplifier
always matches the ratio obtained from Scovil and Schulz-DuBois's
intuitive definition. Before we calculate the engine's efficiency,
we wish to obtain Carnot's formulation of the second law in
differential form. We begin with Spohn's entropy production
function:
\begin{equation}
\sigma^{SC}=\frac{\partial S}{\partial
t}-\frac{\dot{Q}_{H}}{T_{H}}-\frac{\dot{Q}_{C}}{T_{C}}\geq0,\label{Spohn3lsc}
\end{equation}
where $\frac{\partial S}{\partial t}$ is the three-level system
entropy change, and $-\frac{\dot{Q}_{H(C)}}{T_{H(C)}}$ is the
entropy flux from/to the hot (cold) reservoir. At steady state
$\frac{\partial S}{\partial t}=0$. We now wish to rewrite
$\dot{Q}_{C}^{ss}$ in terms of $\dot{Q}_{H}^{ss}$ and $P^{ss}$. The
quantity $\dot{E}\equiv\textrm{Tr}\{\dot{\rho H}\}$, which measures
the energy flux including the atomic-field interaction energy is
given by:
\begin{equation}
\dot{E}=\textrm{Tr}\left\{\frac{\partial\rho}{\partial
t}H\right\}+\textrm{Tr}\left\{\rho \frac{\partial H}{\partial
t}\right\}=\dot{Q}_{H}+\dot{Q}_{C}+P.
\end{equation}
At steady state:
\begin{equation}
\dot{E}^{ss}=\dot{Q}_{H}^{ss}+\dot{Q}_{C}^{ss}+P^{ss}=\frac{2\Gamma_{01}\Gamma_{02}
\lambda_{sc}^3(n_{01}-n_{02})[\omega-(\omega_{1}-\omega_{2})]}{F}.\label{Edot3lscss}
\end{equation}
The quantity $\dot{E}^{ss}$ is zero only at perfect atomic-field
resonance. However, the quantity
$\dot{E}_{m}\equiv\textrm{Tr}\{\dot{\rho}H_{m}\}$, which measures
the energy flux without the atomic-field interaction energy, and was
introduced originally in \cite{Erez02}, is zero at steady state, as
$H_{m}$ does not depend on time. Expanding $\dot{E}_{m}$ yields:
\begin{equation}
\dot{E}_{m}\equiv\textrm{Tr}\{\dot{\rho}H_{m}\}=\dot{Q}_{Hm}+\dot{Q}_{Cm}+P_{m},
\end{equation}
where
$\dot{Q}_{H(C)m}=\textrm{Tr}\{\mathcal{L}_{dH(C)}[\rho]H_{m}\}$ and
$P_{m}=-\frac{i}{\hbar}\textrm{Tr}\{\rho[H_{m},V(t)]\}$ are the
alternative definitions for heat flux and power introduced in
\cite{Erez02}. At steady state: (1) $\dot{E}_{m}=0$ and hence
$\dot{Q}_{Cm}^{ss}=-(\dot{Q}_{Hm}^{ss}+P_{m}^{ss})$, and (2) since
$\textrm{Tr}\{\mathcal{L}_{dH(C)}[\rho^{ss}]V\}=0$,
$\dot{Q}_{H(C)m}^{ss}=\dot{Q}_{H(C)}^{ss}$. Therefore we can replace
$\dot{Q}_{C}^{ss}$ in equation \ref{Spohn3lsc} with
$-(\dot{Q}_{H(C)}^{ss}+P_{m}^{ss})$ where
\begin{equation}
P_{m}^{ss}=-\frac{2\Gamma_{01}\Gamma_{02}\lambda_{sc}^3(n_{01}-n_{02})(\omega_{1}-\omega_{2})}{F},
\end{equation}
and obtain:
\begin{equation}
\eta\equiv-\frac{P_{m}^{ss}}{\dot{Q}_{H}^{ss}}\leq\frac{T_{H}-T_{C}}{T_{H}},\label{Carnot3lsc}
\end{equation}
which is Carnot's efficiency formula in differential form. We note
that equation \ref{Carnot3lsc} is always true regardless of a
resonance condition. Moreover, we wish to emphasize that rewriting
$\dot{Q}_{C}^{ss}$ in terms of $\dot{Q}_{H}^{ss}$ and $P_{m}^{ss}$
for non-resonant cases is possible only through the alternative
approach to energy flux in unipartite systems discussed in
\cite{Erez02}.

Substitution of $P_{m}^{ss}$ (which is identical with $P^{ss}$ at
perfect resonance) and $\dot{Q}_{H}^{ss}$ into the engine's
efficiency formula at steady state ($\frac{\partial S_{m}}{\partial
t}=0$) yields:
\begin{equation}
\eta=-\frac{P_{m}^{ss}}{\dot{Q}_{H}^{ss}}=\frac{\omega_{1}-\omega_{2}}{\omega_{1}-\omega_{0}}=\frac{\omega_{s}}{\omega_{p}},
\end{equation}
which is identical with the maser's efficiency defined intuitively
by Scovil and Schulz-DuBois. Our model offers a statistical
description for the reservoirs, and it allows us to derive
thermodynamic fluxes, which in turn yield Scovil and Schulz-DuBois's
efficiency formula.

We note that Geva and Kosloff \cite{Geva04} also considered a
semiclassical model for a three-level amplifier. The main difference
between their model and the semiclassical ED-JCM is that in Geva and
Kosloff's model the time dependence of the classical field affects
the dissipative superoperator. As a result, the steady state
efficiency in the model by Geva and Kosloff depends on the power of
the field, and hence it is generally not the same as in Scovil and
Schulz-DuBois's intuitive definition.

\subsection{Steady state inversion ratio}

In their early work \cite{Scovil01} Scovil and Schulz-DuBois
asserted that the ground state population ($\rho_{00}$) is bigger
than the populations in the two excited states ($\rho_{11}$ and
$\rho_{22}$). This is indeed verified in Appendix II. They also
asserted that the inversion ratio between the two excited levels is
given by:
\begin{equation}
r\equiv\frac{\rho_{11}}{\rho_{22}}=e^{-\frac{\hbar(E_{1}-E_{0})}{k_{B}T_{H}}}e^{\frac{\hbar(E_{0}-E_{2})}{k_{B}T_{C}}}=e^{-\frac{\hbar(\omega_{1}-\omega_{0})}{k_{B}T_{H}}}e^{\frac{\hbar(\omega_{0}-\omega_{2})}{k_{B}T_{C}}}\label{Scovilwrong}.
\end{equation}
This assertion appears to be well motivated physically, since it
would seem that at steady state the population ratios between the
two excited levels and the pumping level should be related by
Boltzmann factors. However, it turns out that this is not correct.
While it is true that the matter reaches a steady state, as seen in
both the quantum and semiclassical models, the matter-field system
as a whole does not reach a steady state, as was seen by solving the
fully quantum model in this paper. Moreover, there is no a priori
requirement of what steady state populations will be attained. We
will now demonstrate that the ratio between the two excited levels
asserted by Scovil and Schulz-DuBois is not correct. Substituting
the expressions for the reservoirs' temperatures given in equation
\ref{restemp} into equation \ref{Scovilwrong} yields:
\begin{equation}
r=e^{\ln(1/n_{02}+1)}e^{-\ln(1/n_{01}+1)}=\frac{n_{01}(n_{02}+1)}{n_{02}(n_{01}+1)}\label{Scovilwrongsub}.
\end{equation}
Substituting $n_{01}=10,n_{02}=0.1$ yields $r=10$. In Appendix II we
give analytical expressions for all the density matrix elements at
steady state, from which a closed formula for $r$ may be obtained:
\begin{equation}
r=r(\Gamma_{01},\Gamma_{02},n_{01},n_{02},\lambda_{sc})=\frac{B}{C},
\end{equation}
where $B,C$ are positive constants given in Appendix II.
Substituting $n_{01}=10,n_{02}=0.1,
\Gamma_{01}=\Gamma_{02}=\Gamma=0.001$ in the analytical expression
for $r$ yields (similarly to the quantum model) only a marginal
inversion ratio between the two excited levels,
$r=\{1.01,1.00001,1.00000001\}$ for field strengths
$E_{0}=\{0.1,1,10\}$, respectively, where equation
\ref{Scovilwrongsub} yields $r=10$.

\section{Conclusion}

We have analyzed a fully quantum model in which a three-level
material system is coupled to a single quantized cavity mode and two
thermal photonic reservoirs in a framework of a heat engine. This
gives what is arguably the the simplest possible quantum model for
light amplification.  At the same time, it permits a full
thermodynamic analysis.  Unlike previous work \cite{Geva04}, the
field is not considered as an external time dependent force acting
on the matter, but it is an integral part of the quantum system,
allowing us to treat both light and matter on equal footing. We
solved the ED-JCM master equation numerically, and showed that
indeed amplification of the selected cavity mode occurs even in this
simple model. However, initial field coherence is lost, as seen by
the radially symmetric $Q$ function for $t\geq(2\Gamma)^{-1}$.
Moreover, we find that the quantized field mode has an entropy
content that changes dramatically at short times, and increases very
slowly for $t\geq(2\Gamma)^{-1}$. The matter-field system as a whole
never reaches a steady state: at $t\geq(2\Gamma)^{-1}$ the energy in
the field continues to increase linearly in time, which can be
analyzed thermodynamically in terms of power generation from energy
in the hot reservoir. The three-level matter system, obtained by
performing the partial trace of the full system over the field, does
reach a steady state as seen by constant steady state energy and
entropy.

Another aspect of the quantum treatment that cannot be dealt with at
all within the framework of the semiclassical ED-JCM is
entanglement. We showed that at short times $t<(2\Gamma)^{-1}$ the
matter-field system is entangled, as seen by the negative
conditional entropies and the negativity of the partially transposed
density matrix. However, at longer times $t>(2\Gamma)^{-1}$ we
believe that the matter-field system is classically correlated but
not entangled, as the conditional entropies (which are almost equal
to the partial entropies) and the partially transposed density
matrix are both positive.

Based on our previous work on bipartite systems governed by a time
independent master equation \cite{Erez02} we were able to derive the
fundamental laws of thermodynamics. The first law is obtained both
for the full matter-field system and for the individual (partially
traced) subsystems, using thermodynamical fluxes of heat flux and
power. The second law of thermodynamics in differential form is
guaranteed to exist for the full matter-field system through Spohn's
\cite{Spohn} entropy production function. We define a new entropy
production function $\sigma_m$ based on matter thermodynamical
fluxes. Through $\sigma_m$ we show that at steady state, when the
main entropy production is due to heat fluxes from/to the heat
reservoirs, Carnot's efficiency formula is obtained in differential
form.

A strong motivation for this work comes from an early paper by
Scovil and Schulz-DuBois \cite{Scovil01} in which they analyze a
three-level maser as a heat engine. In their work, they intuitively
defined the engine's efficiency as the ratio between the maser
frequency and the pumping frequency. However, they do not connect
this efficiency with explicit expressions for work and heat, as
expected from a thermodynamical analysis of a heat engine. In our
quantized field treatment, the efficiency formula of Scovil and
Schulz-DuBois was found to be in complete agreement with numerical
calculations based on thermodynamical power and heat fluxes.

We have also analyzed a semiclassical version of the ED-JCM. We
obtained closed analytical expressions for power and heat flux at
steady state that are in virtually perfect agreement with those
obtained numerically for the fully quantum ED-JCM. One may conclude
from this that as far as steady state thermodynamical fluxes are
concerned, the semiclassical model is sufficient. Furthermore, from
our analytical results for power and heat flux we were able to
recover Scovil and Schulz-DuBois's efficiency formula analytically.
One of the assertions in the work of Scovil and Schulz-DuBois is
that the ratio of populations in the two excited levels is given by
a product of Boltzmann factors. We showed analytically that this
last assertion does not hold in general.

In future work, we intend to explore further the other thermodynamic
scenarios implied by the present model, both semiclassically and
quantum mechanically.  Of particular interest is the reversal of the
present mode of operation of the engine so that it operates as a
refrigerator for light.

\section*{Acknowledgments}
This work was supported by the German-Israeli Foundation for
Scientific Research and Development.

\section*{Appendix I: Derivation of the damped JCM master equation}

The master equation for the resonant Jaynes-Cummings model (JCM)
with atomic damping in the interaction representation is given by:
\begin{equation}
\mbox{\boldmath$\dot{\rho}_{af}^{I}$}=\mathcal{L}_{h}[\mbox{\boldmath$\rho_{af}^{I}$}]+\mathcal{L}_{d}[\mbox{\boldmath$\rho_{af}^{I}$}],\label{JCMdafI1}
\end{equation}
where $\mbox{\boldmath$\rho_{af}^{I}$}$ is the combined atom-field
density matrix in the interaction picture, and
$\mathcal{L}_{h}[\mbox{\boldmath$\rho_{af}^{I}$}]$ and
$\mathcal{L}_{d}[\mbox{\boldmath$\rho_{af}^{I}$}]$ are given by:
\begin{eqnarray}
\mathcal{L}_{h}[\mbox{\boldmath$\rho_{af}^{I}$}]&=&-\frac{i}{\hbar}[\mbox{\boldmath$V^{I}$},\mbox{\boldmath$\rho^{I}_{af}$}]=-i[\lambda(\mbox{\boldmath$\sigma^{-}a^{\dag}$}+\mbox{\boldmath$\sigma^{+}a$}),\mbox{\boldmath$\rho^{I}_{af}$}]\nonumber\\
\mathcal{L}_{d}[\mbox{\boldmath$\rho_{af}^{I}$}]&=&
\!\!\Gamma(n_{th}\!+\!1)([\mbox{\boldmath$\sigma^{-}$},\mbox{\boldmath$\rho^{I}_{af}\sigma^{+}$}]\!\!+\!\![\mbox{\boldmath$\sigma^{-}\rho^{I}_{af}$},\mbox{\boldmath$\sigma^{+}$}])\!\!+
\!\!\Gamma
n_{th}\!([\mbox{\boldmath$\sigma^{+}$},\mbox{\boldmath$\rho^{I}_{af}\sigma^{-}$}]\!+\![\mbox{\boldmath$\sigma^{+}\rho^{I}_{af}$},\mbox{\boldmath$\sigma^{-}$}]),\label{JCMdafI2}
\end{eqnarray}
where $\mbox{\boldmath$\sigma^{+}(a^{\dag}$})$ and
$\mbox{\boldmath$\sigma^{-}(a)$}$ are atomic (field) creation and
annihilation operators
($\mbox{\boldmath$\sigma^{+}\sigma^{-}$}-\mbox{\boldmath$\sigma^{-}\sigma^{+}$}=\mbox{\boldmath$\sigma_{z}$}$;
$\mbox{\boldmath$\sigma_{z}$}$ being the Pauli $z$ matrix).
$\lambda$, $\Gamma$, $n_{th}$ are the atomic-field coupling
constant, Weiskopf-Wigner decay constant, and the number of thermal
photons, respectively. Note that since equation \ref{JCMdafI1} is a
master equation of a bipartite system, all the operators in equation
\ref{JCMdafI2} are implicitly tensor products. For example,
$\mbox{\boldmath$\sigma^{+}\sigma^{-}\rho^{I}_{af}$}$ is shorthand
notation for
$(\sigma_{+}\otimes\openone_{f})(\sigma_{-}\otimes\openone_{f})\mbox{\boldmath$\rho^{I}_{af}$}$.
The damped JCM master equation is usually obtained by adding the
Hamiltonian part and the dissipative part. We note that van Wonderen
gave an analytical solution for the atomic density matrix in the
damped JCM \cite{vanWonderen01}, and later studied the entropic
behavior of the atom \cite{vanWonderen02}. In this appendix, we
derive the full JCM master equation by applying the weak-coupling,
Markovian and Weiskopf-Wigner approximations, and using a set of
unitary transformations. The derivation of the dissipative part
follows closely the derivation given by Scully and Zubairy
\cite{S&Z}.

We start with the full system (atom-field)-bath Hamiltonian in the
Schr\"{o}dinger picture:
\begin{equation}
\mbox{\boldmath$\hat{H}$}=\mbox{\boldmath$\hat{H}_{s}$}+\mbox{\boldmath$\hat{H}_{b}$}+\mbox{\boldmath$\hat{V}_{sb}$},
\end{equation}
where $\mbox{\boldmath$\hat{H}_{s}$}$,
$\mbox{\boldmath$\hat{H}_{b}$}$, $\mbox{\boldmath$\hat{V}_{sb}$}$
are given by:
\begin{eqnarray}
\mbox{\boldmath$\hat{H}_{s}$}&=&\mbox{\boldmath$\hat{H}_{a}$}+\mbox{\boldmath$\hat{H}_{f}$}+\mbox{\boldmath$\hat{V}_{af}$}=\hbar\frac{\omega_{a}}{2}\mbox{\boldmath$\hat{\sigma}_{z}$}+\hbar\omega_{f}
\mbox{\boldmath$\hat{a}^{\dag}\hat{a}$}
+\hbar\lambda(\mbox{\boldmath$\hat{\sigma}^{-}\hat{a}^{\dag}$}+\mbox{\boldmath$\hat{\sigma}^{+}\hat{a}$})\nonumber\\
\mbox{\boldmath$\hat{H}_{b}$}&=&\hbar\sum_{k}{\omega_{k}\mbox{\boldmath$\hat{a}_{k}^{\dag}\hat{a}_{k}$}}\nonumber\\
\mbox{\boldmath$\hat{V}_{sb}$}&=&\hbar\sum_{k}{\lambda_{k}(\mbox{\boldmath$\hat{\sigma}^{-}\hat{a}_{k}^{\dag}$}+\mbox{\boldmath$\hat{\sigma}^{+}\hat{a}_{k}$})}.
\end{eqnarray}
We denote by $s$ the atom-field system, by $b$ the bath which is
composed of an infinite number of oscillators where the operators of
each oscillator are denoted by subscript $k$, and by $\lambda_{k}$
the atomic-$k$th mode coupling constant. The hat notation indicates
that all operators are implicitly tensor products with the
appropriate identity operators. For example
$\mbox{\boldmath$\hat{\sigma}^{-}$}=\sigma^{-}\otimes\openone_{f}\otimes\openone_{b}$,
$\mbox{\boldmath$\hat{a}$}=\openone_{a}\otimes
a\otimes\openone_{b}$, and
$\mbox{\boldmath$\hat{a}_{k}$}=\openone_{a}\otimes\openone_{f}\otimes
a_{k}$. The above Hamiltonian is written under the rotating wave
approximation (RWA) meaning that only energy conserving terms are
considered. Note that only the atom is coupled directly to the bath
modes. The evolution of the full system-bath is purely Hamiltonian:
\begin{equation}
\mbox{\boldmath$\dot{\rho}_{sb}$}=\mathcal{L}_{h}=-\frac{i}{\hbar}[\mbox{\boldmath$\hat{H}$},\mbox{\boldmath$\rho_{sb}$}].
\end{equation}
We now move to the system-bath interaction picture (denoted by
superscript $\bar{I}$):
\begin{eqnarray}
\mbox{\boldmath$\rho_{sb}^{\bar{I}}$}&=&e^{\frac{i}{\hbar}\mbox{\boldmath$\hat{\bar{H}}_{0}$}t}\mbox{\boldmath$\rho_{sb}$}e^{-\frac{i}{\hbar}\mbox{\boldmath$\hat{\bar{H}}_{0}$}t}\nonumber\\
\mbox{\boldmath$\dot{\rho}_{sb}^{\bar{I}}$}&=&-\frac{i}{\hbar}[\mbox{\boldmath$\hat{V}_{sb}^{\bar{I}}$},\mbox{\boldmath$\rho_{sb}^{\bar{I}}$}],\label{JCMdsbI}
\end{eqnarray}
where:
\begin{eqnarray}
\mbox{\boldmath$\hat{\bar{H}}_{0}$}&=&\mbox{\boldmath$\hat{H}_{a}$}+\mbox{\boldmath$\hat{H}_{f}$}+\mbox{\boldmath$\hat{V}_{af}$}+\mbox{\boldmath$\hat{H}_{b}$}\nonumber\\
\mbox{\boldmath$\hat{V}_{sb}^{\bar{I}}$}&=&
\hbar\sum_{k}{\lambda_{k}[\mbox{\boldmath$\hat{\bar{\sigma}}^{-}(t)a_{k}^{\dag}$}e^{i\omega_{k}t}+\mbox{\boldmath$\hat{\bar{\sigma}}^{+}(t)a_{k}$}e^{-i\omega_{k}t}]}
,
\end{eqnarray}
where
$\mbox{\boldmath$\hat{\bar{\sigma}}^{-(+)}(t)$}=e^{\frac{i}{\hbar}\mbox{\boldmath$\hat{H}_{s}$}t}\mbox{\boldmath$\hat{\sigma}^{-(+)}$}e^{-\frac{i}{\hbar}\mbox{\boldmath$\hat{H}_{s}$}t}$.
In the derivation of equation \ref{JCMdsbI} we made use of the
identity
$[\mbox{\boldmath$\hat{H}_{s}$},\mbox{\boldmath$\hat{H}_{b}$}]=0$. A
perturbation expansion to second order in
$\mbox{\boldmath$\hat{V}_{sb}$}$ yields:
\begin{equation}
\mbox{\boldmath$\dot{\rho}_{sb}^{\bar{I}}$}=-\frac{i}{\hbar}[\mbox{\boldmath$\hat{V}_{sb}^{\bar{I}}(t)$},\mbox{\boldmath$\rho_{sb}^{\bar{I}}(0)$}]
-\frac{1}{\hbar^{2}}\int_{0}^{t}{dt'[\mbox{\boldmath$\hat{V}_{sb}^{\bar{I}}(t)$},[\mbox{\boldmath$\hat{V}_{sb}^{\bar{I}}(t')$},\mbox{\boldmath$\rho_{sb}^{\bar{I}}(t')$}]]}.
\end{equation}
Consider the weak system-bath coupling limit, that is
$\mbox{\boldmath$\rho_{sb}(t)$}=\mbox{\boldmath$\rho_{s}(t)$}\otimes\mbox{\boldmath$\rho_{b}(0)$}+\mbox{\boldmath$\rho_{c}$}$,
where $\mbox{\boldmath$\rho_{c}$}$ is any correlation between the
system and bath which fulfills
$\textrm{Tr}_{b}\{\mbox{\boldmath$\rho_{c}$}\}=0$ (this holds for
$\mbox{\boldmath$\rho_{sb}(t)$}$ both in the Schr\"{o}dinger and
interaction pictures). In this case the atom-field system evolves
according to:
\begin{equation}
\mbox{\boldmath$\dot{\rho}_{s}^{\bar{I}}$}\equiv
\textrm{Tr}_{b}\{\mbox{\boldmath$\rho_{sb}^{\bar{I}}$}\}
=-\frac{i}{\hbar}\textrm{Tr}_{b}\{[\mbox{\boldmath$\hat{V}_{sb}^{\bar{I}}(t)$},\mbox{\boldmath$\rho_{s}^{\bar{I}}(0)$}\otimes\mbox{\boldmath$\rho_{b}(0)$}]\}
-\frac{1}{\hbar^{2}}\textrm{Tr}_{b}\left\{\int_{0}^{t}{dt'[\mbox{\boldmath$\hat{V}_{sb}^{\bar{I}}(t)$},[\mbox{\boldmath$\hat{V}_{sb}^{\bar{I}}(t')$},\mbox{\boldmath$\rho_{s}^{\bar{I}}$}\otimes\mbox{\boldmath$\rho_{b}(0)$}]]}\right\}.
\label{rhosnon}
\end{equation}
Note that in equation \ref{rhosnon},
$\mbox{\boldmath$\rho_{b}(0)$}=\mbox{\boldmath$\rho_{b}^{\bar{I}}(0)$}$
and
$\mbox{\boldmath$\rho_{s}^{\bar{I}}$}=e^{\frac{i}{\hbar}\mbox{\boldmath$\hat{H}_{s}$}t}\mbox{\boldmath$\rho_{s}$}e^{-\frac{i}{\hbar}\mbox{\boldmath$\hat{H}_{s}$}t}$.
The explicit form of equation \ref{rhosnon} is given by:
\begin{eqnarray}
\mbox{\boldmath$\dot{\rho}_{s}^{\bar{I}}$}&=&-i\sum_{k}{\lambda_{k}\langle
\mbox{\boldmath$\hat{a}_{k}^{\dag}$}\rangle[\mbox{\boldmath$\hat{\bar{\sigma}}^{-}(t)$},\mbox{\boldmath$\rho_{s}^{\bar{I}}(0)$}]e^{i\omega_{k}t}}\nonumber\\
&-&\int_{0}^{t}dt'\sum_{k,k'}\lambda_{k}\lambda_{k'}
\left\{[\mbox{\boldmath$\hat{\bar{\sigma}}^{-}(t)\hat{\bar{\sigma}}^{+}(t')\rho_{s}^{\bar{I}}(t')$}-\mbox{\boldmath$\hat{\bar{\sigma}}^{+}(t')\rho_{s}^{\bar{I}}(t')\bar{\sigma}_{-}(t)$}]
e^{i\omega_{k}t-i\omega_{k'}t'}\langle
\mbox{\boldmath$a_{k}^{\dag}a_{k'}$}\rangle\right.\nonumber\\
&+&[\mbox{\boldmath$\hat{\bar{\sigma}}^{+}(t)\hat{\bar{\sigma}}^{-}(t')\rho_{s}^{\bar{I}}(t')$}-\mbox{\boldmath$\hat{\bar{\sigma}}^{-}(t')\rho_{s}^{\bar{I}}(t')\bar{\sigma}^{+}(t)$}]
e^{-i\omega_{k}t+i\omega_{k'}t'}\langle \mbox{\boldmath$a_{k}a_{k'}^{\dag}$}\rangle\nonumber\\
&+&\left.[\mbox{\boldmath$\hat{\bar{\sigma}}^{-}(t)\hat{\bar{\sigma}}^{-}(t')\rho_{s}^{\bar{I}}(t')$}
-\mbox{\boldmath$\hat{\bar{\sigma}}^{-}(t')\rho_{s}^{\bar{I}}(t')\hat{\bar{\sigma}}^{-}(t)$}
-\mbox{\boldmath$\hat{\bar{\sigma}}^{-}(t)\rho_{s}^{\bar{I}}(t')\hat{\bar{\sigma}}^{-}(t')$}\right.\nonumber\\
&+&\left.\mbox{\boldmath$\rho_{s}^{\bar{I}}(t')\hat{\bar{\sigma}}^{-}(t')\hat{\bar{\sigma}}^{-}(t)$}]e^{i\omega_{k}t-i\omega_{k'}t'}\langle
\mbox{\boldmath$a_{k}^{\dag}a_{k'}^{\dag}$}\rangle\right\}+\textrm{H.c.},\label{rhosexp0}
\end{eqnarray}
where H.c. refers to all terms on the RHS. The bath density matrix
is now assumed to be composed of a product of oscillatory modes each
being in a thermal state, that is:
\begin{equation}
\mbox{\boldmath$\rho_{b}$}=\prod_{k}{\rho_{k}};\
\rho_{k}=\sum_{n_{k}}\frac{\bar{n}_{k}^{n_{k}}}{(\bar{n}_{k}+1)^{(n_{k}+1)}}|{n_{k}}\rangle\langle
{n_{k}}|,
\end{equation}
where $\bar{n}_{k}$ is the average number of thermal photons in the
$k$th mode. With this assumption equation \ref{rhosexp0} reduces to:
\begin{eqnarray}
\mbox{\boldmath$\dot{\rho}_{s}^{\bar{I}}$}&=&-\int_{0}^{t}dt'\sum_{k}\lambda_{k}^{2}
\left\{[\mbox{\boldmath$\bar{\sigma}^{-}(t)\bar{\sigma}^{+}(t')\rho_{s}^{\bar{I}}(t')$}\}-\mbox{\boldmath$\bar{\sigma}^{+}(t')\rho_{s}^{\bar{I}}(t')\bar{\sigma}^{-}(t)$}\}]
e^{i\omega_{k}(t-t')}\bar{n}_{k}\right.\\
&+&\left.[\mbox{\boldmath$\bar{\sigma}^{+}(t)\bar{\sigma}^{-}(t')\rho_{s}^{\bar{I}}(t')$}-\mbox{\boldmath$\bar{\sigma}^{-}(t')\rho_{s}^{\bar{I}}(t')\bar{\sigma}^{+}(t)$}]
e^{-i\omega_{1}(t-t')}(\bar{n}_{k}+1)\right\}+\textrm{H.c.}\label{rhosexp1}
\end{eqnarray}
The sum over $k$ is now replaced by an integral:
\begin{equation*}
\sum_{k}\rightarrow
2\frac{V}{(2\pi)^{3}}\int_{0}^{2\pi}d\phi\int_{0}^{\pi}d\theta\sin{\theta}\int_{0}^{\infty}d\nu_{k}\frac{\nu_{k}^{2}}{c^{3}},
\end{equation*}
where $V$ is the quantization volume and $\nu_{k}$ is the $k$th mode
oscillation frequency. Substituting
$\lambda_{k}^{2}=\frac{\nu_{k}}{2\hbar\varepsilon_{0}V}\mathcal{D}^{2}\cos^{2}{\theta}$
($\mathcal{D}$ is the transition dipole matrix element, and $\theta$
is the angle between $\mathcal{D}$ and the electric field
polarization vector), and integrating in the Weiskopf-Wigner
approximation (extending the lower limit of the integral over
$\nu_{k}$ from $0$ to $-\infty$, and replacing
$\nu_{k}=2\pi\omega_{k}$ by $\omega$) simplifies equation
\ref{rhosexp1}:
\begin{eqnarray}
\mbox{\boldmath$\dot{\rho}_{s}^{\bar{I}}$}&=&\mathcal{\bar{L}}_{d}[\mbox{\boldmath$\rho_{s}^{\bar{I}}$}]=-\Gamma
n_{th}
[\mbox{\boldmath$\bar{\sigma}^{-}(t)\bar{\sigma}^{+}(t)\rho_{s}^{\bar{I}}(t)$}-\mbox{\boldmath$\bar{\sigma}^{+}(t)\rho_{s}^{\bar{I}}(t)\bar{\sigma}^{-}(t)$}]\nonumber\\
&-&\Gamma(n_{th}+1)[\mbox{\boldmath$\bar{\sigma}^{+}(t)\bar{\sigma}^{-}(t)\rho_{s}^{\bar{I}}(t)$}-\mbox{\boldmath$\bar{\sigma}^{-}(t)\rho_{s}^{\bar{I}}(t)\bar{\sigma}^{+}(t)$}]
+\textrm{H.c.},\label{rhosexp2}
\end{eqnarray}
where $n_{th}\equiv\bar{n}_{k_{0}}$ $(k_{0}=\omega/c)$ is the
average number of thermal photons, and
$\Gamma=\frac{\omega^{3}\mathcal{D}^{2}}{6\pi\hbar\varepsilon_{0}c^{3}}$
is the decay rate.

We now move to the system Schr\"{o}dinger picture:
\begin{eqnarray}
\mbox{\boldmath$\rho_{s}$}&=&\mbox{\boldmath$e^{-\frac{i}{\hbar}H_{s}t}\rho_{s}^{\bar{I}}e^{\frac{i}{\hbar}H_{s}t}$}\nonumber\\
\mbox{\boldmath$\dot{\rho}_{s}$}&=&-\frac{i}{\hbar}[\mbox{\boldmath$H_{s}$},\mbox{\boldmath$\rho_{s}$}]+\mbox{\boldmath$e^{-\frac{i}{\hbar}H_{s}t}\mathcal{\bar{L}}_{d}[\rho_{s}^{\bar{I}}]e^{\frac{i}{\hbar}H_{s}t}$}.\label{JCMdsS}
\end{eqnarray}
Using the definitions for $\mbox{\boldmath$\bar{\sigma}^{-(+)}(t)$}$
and $\mbox{\boldmath$\rho_{s}^{\bar{I}}$}$ (after tracing out the
bath) it is easily shown that
$\mbox{\boldmath$e^{-\frac{i}{\hbar}H_{s}t}$}\mathcal{\bar{L}}_{d}[\mbox{\boldmath$\rho_{s}^{\bar{I}}$}]\mbox{\boldmath$e^{\frac{i}{\hbar}H_{s}t}$}=\mathcal{L}_{d}[\mbox{\boldmath$\rho_{s}$}]$.
Finally, the master equation for the system in the Schr\"{o}dinger
picture is given by:
\begin{eqnarray}
\mbox{\boldmath$\dot{\rho}_{s}$}&=&\mathcal{L}_{h}[\mbox{\boldmath$\rho_{s}$}]+\mathcal{L}_{d}[\mbox{\boldmath$\rho_{s}$}]\nonumber\\
\mathcal{L}_{h}[\mbox{\boldmath$\rho_{s}$}]&=&-\frac{i}{\hbar}[\mbox{\boldmath$H_{s}$},\mbox{\boldmath$\rho_{s}$}]\nonumber\\
\mathcal{L}_{d}[\mbox{\boldmath$\rho_{s}$}]&=&
\Gamma(n_{th}+1)([\mbox{\boldmath$\sigma^{-}$},\mbox{\boldmath$\rho_{s}\sigma_{+}$}]+[\mbox{\boldmath$\sigma^{-}\rho_{s}$},\mbox{\boldmath$\sigma^{+}$}])+
\Gamma
n_{th}([\mbox{\boldmath$\sigma^{+}$},\mbox{\boldmath$\rho_{s}\sigma^{-}$}]+[\mbox{\boldmath$\sigma^{+}\rho_{s},\sigma_{-}$}]),\label{rhosexpS}
\end{eqnarray}
where we deliberately omitted the superscript $S$ labeling the
Schr\"{o}dinger picture.

To summarize, we went through the following path:
\begin{equation*}
\mbox{\boldmath$\rho_{sb}$}\rightarrow\mbox{\boldmath$\rho_{sb}^{\bar{I}}$}\rightarrow\mbox{\boldmath$\rho_{s}^{\bar{I}}$}\rightarrow\mbox{\boldmath$\rho_{s}$}.
\end{equation*}
The first transition takes us from the system-bath Schr\"{o}dinger
picture to the system-bath interaction picture through a unitary
transformation. Tracing over the bath under the weak coupling,
Markovian, and Weiskopf-Wigner approximations leads us to the system
dissipative interaction picture. Finally, by applying a unitary
transformation we move to the system Schr\"{o}dinger picture.

To complete the analysis we now move to the standard interaction
picture which includes both the Hamiltonian and the dissipative
parts:
\begin{eqnarray}
\mbox{\boldmath$\rho_{s}^{I}$}&=&\mbox{\boldmath$e^{\frac{i}{\hbar}(H_{a}+H_{f})t}\rho_{s}e^{\frac{i}{\hbar}(H_{a}t+H_{f})t}$}\nonumber\\
\mbox{\boldmath$\dot{\rho}_{s}^{I}$}&=&-\frac{i}{\hbar}[\mbox{\boldmath$V$},\mbox{\boldmath$\rho_{s}^{I}$}]+\mbox{\boldmath$e^{\frac{i}{\hbar}(H_{a}+H_{f})t}$}\mathcal{L}_{d}[\mbox{\boldmath$\rho_{s}$}]\mbox{\boldmath$e^{\frac{i}{\hbar}(H_{a}+H_{f})t}$},\label{JCMdsI}
\end{eqnarray}
where it can be shown that:
$\mbox{\boldmath$e^{-\frac{i}{\hbar}(H_{a}+H_{f})t}$}\mathcal{L}_{d}[\mbox{\boldmath$\rho_{s}$}]\mbox{\boldmath$e^{\frac{i}{\hbar}(H_{a}+H_{f})t}$}=\mathcal{L}_{d}[\mbox{\boldmath$\rho_{s}^{I}$}]$.
Equation \ref{JCMdsI} is now identical with equation \ref{JCMdafI2},
with subscript $s$ replacing subscript $af$.

\section*{Appendix II: Density matrix of the semiclassical ED-JCM amplifier at steady state}

The density matrix for the semiclassical ED-JCM operating as an
amplifier ($\rho_{11}-\rho_{22}>0$) is given by:
\begin{equation}
\rho^{ss}=\left(
\begin{array}{ccc}
  \rho_{00} & 0 & 0 \\
  0 & \rho_{11} & i|\rho_{12}| \\
  0 & -i|\rho_{12}| & \rho_{22} \\
\end{array}
\right)=\left(
\begin{array}{ccc}
  A/F & 0 & 0 \\
  0 & B/F & iD/F \\
  0 & -iD/F & C/F  \\
\end{array}
\right),
\end{equation}
where $A,B,C,D,F$ are given by:
\begin{eqnarray}
A&=&\lambda_{sc}^{3}\Gamma_{02}+\lambda_{sc}^{3}\Gamma_{01}+\lambda_{sc}^{3}\Gamma_{02}n_{02}+\lambda_{sc}^{3}\Gamma_{01}n_{01}+\Gamma_{02}\Gamma_{01}^{2}+\Gamma_{02}\Gamma_{01}^{2}n_{02}+2\Gamma_{02}\Gamma_{01}^{2}n_{01}+\Gamma_{02}\Gamma_{01}^{2}n_{01}^{2}\nonumber\\
\
&+&2\Gamma_{02}\Gamma_{01}^2n_{02}n_{01}+\Gamma_{02}\Gamma_{01}^2n_{02}n_{01}^{2}+\Gamma_{02}^{2}\Gamma_{01}+2\Gamma_{02}^{2}\Gamma_{01}n_{02}+2\Gamma_{02}^{2}\Gamma_{01}n_{02}+\Gamma_{02}^{2}\Gamma_{01}n_{01}\nonumber\\
\ &+&2\Gamma_{02}^{2}\Gamma_{01}n_{02}n_{01}+\Gamma_{02}^{2}\Gamma_{01}n_{02}^{2}n_{01}\nonumber\\
B&=&\lambda_{sc}^{3}\Gamma_{02}n_{02}+\lambda_{sc}^{3}\Gamma_{01}n_{01}+\Gamma_{02}\Gamma_{01}^{2}n_{01}+\Gamma_{02}\Gamma_{01}^{2}n_{01}^{2}+\Gamma_{02}\Gamma_{01}^{2}n_{02}n_{01}+\Gamma_{02}\Gamma_{01}^{2}n_{02}n_{01}^{2}\nonumber\\
\
&+&\Gamma_{02}^{2}\Gamma_{01}n_{01}+2\Gamma_{02}^{2}\Gamma_{01}n_{02}n_{01}+\Gamma_{02}^{2}\Gamma_{01}n_{02}^{2}n_{01}\nonumber\\
C&=&\lambda_{sc}^{3}\Gamma_{02}n_{02}+\lambda_{sc}^{3}\Gamma_{01}n_{01}+\Gamma_{02}\Gamma_{01}^{2}n_{02}+2\Gamma_{02}\Gamma_{01}^{2}n_{02}n_{01}+\Gamma_{02}\Gamma_{01}^{2}n_{02}n_{01}^{2}\nonumber\\
\
&+&\Gamma_{02}^{2}\Gamma_{01}n_{02}+\Gamma_{02}^{2}\Gamma_{01}n_{02}^{2}+\Gamma_{02}^{2}\Gamma_{01}n_{02}n_{01}+\Gamma_{02}^{2}\Gamma_{01}n_{02}^{2}n_{01}\nonumber\\
D&=&\lambda_{sc}^{2}\Gamma_{02}\Gamma_{01}(n_{01}-n_{02})\nonumber\\
F&=&\lambda_{sc}^{3}\Gamma_{02}+\lambda_{sc}^{3}\Gamma_{01}+3\lambda_{sc}^{3}\Gamma_{02}n_{02}+3\lambda_{sc}^{3}\Gamma_{01}n_{01}+\Gamma_{02}\Gamma_{01}^{2}+3\Gamma_{02}\Gamma_{01}^{2}n_{01}+2\Gamma_{02}\Gamma_{01}^{2}n_{01}^{2}+5\Gamma_{02}\Gamma_{01}^{2}n_{02}n_{01}\nonumber\\
\
&+&3\Gamma_{02}\Gamma_{01}^{2}n_{02}n_{01}^{2}+\Gamma_{02}^{2}\Gamma_{01}+3\Gamma_{02}^{2}\Gamma_{01}n_{02}+2\Gamma_{02}^{2}\Gamma_{01}n_{02}^{2}+5\Gamma_{02}^{2}\Gamma_{01}n_{02}n_{01}+3\Gamma_{02}^{2}\Gamma_{01}n_{02}^{2}n_{01}.
\end{eqnarray}
$A,B,C,E$ are all positive constants, and since
$A>B,C\Rightarrow\rho_{00}>\rho_{11},\rho_{22}$. Thus the population
in the zeroth (pumping) level is always greater then the population
in either level $|0\rangle$ or $|1\rangle$. Since $D/E=|\rho_{12}|$,
a mathematical feasible expression is obtained only if
$n_{01}>n_{02}$.

\end{document}